\documentclass{tlp}

\jnlPage{1}{?}
\jnlDoiYr{2026}
\doival{10.1017/xxxxx}

\usepackage[utf8]{inputenc}
\usepackage{hyperref}
\usepackage{mathptmx}
\usepackage{amsmath}
\usepackage{amssymb}
\usepackage{stmaryrd} 

\usepackage{xspace}
\usepackage[nameinlink,capitalise,noabbrev]{cleveref}
\usepackage{csquotes}
\usepackage[textfont={small,md}]{caption}
\usepackage[normalem]{ulem}
\usepackage[english]{babel}
\PassOptionsToPackage{babel}{microtype}
\usepackage{microtype}
\usepackage{tabularx}
\usepackage{float}
\PassOptionsToPackage{table,dvipsnames}{xcolor}
\usepackage[dvipsnames]{xcolor}
\usepackage{colortbl}
\usepackage{bookmark}
\usepackage{mathtools}

\newcommand{\codefont}{\small\tt}
\newcommand{\mcode}[1]{\mbox{\codefont{#1}}}
\newcommand{\ccode}[1]{``\mcode{#1}''}
\newcommand{\fcode}[1]{\mbox{\codefont{\footnotesize{#1}}}} 
\newcommand{\us}{\raise-.8ex\hbox{-}}
\newcommand{\funCurry}{\rightarrow_{\texttt{C}}} 
\newcommand{\bind}{\texttt{\normalshape{}>\!>{}=}} 
\newcommand{\listline}{\vrule width0pt depth2.0ex}
\newcommand{\tr}{\mathit{tr}} 
\newcommand{\rename}{\mbox{$\mathit{rename}$}}

\usepackage{listings}
\lstdefinelanguage{haskell}{
  keywords={
    as,case,class,data,default,deriving,do,else,hiding,if,import,in,infix,
    infixl,infixr,instance,let,module,newtype,of,qualified,then,type,where,
  },
  morecomment=[l]--,
  morecomment=[n]{\{-}{-\}},
  morestring=[b]",
  sensitive=true,
}
\newcommand{\haskellann}{\par\vspace*{0.25\baselineskip}\mbox{}{\scriptsize\hfill{\color{gray}\texttt{Haskell}}}\vspace{-1.15\baselineskip}}
\lstnewenvironment{haskell}{\haskellann \lstset{language=haskell}}{}
\newcommand{\haskellinline}[1]{\lstinline[language=haskell,flexiblecolumns=false,mathescape=true,basewidth=0.55em]|#1|}
\lstdefinelanguage{curry}{
  language=haskell,
  morekeywords={free},
}
\newcommand{\curryann}{\par\vspace*{0.25\baselineskip}\mbox{}{\scriptsize\hfill{\color{gray}\texttt{Curry}}}\vspace*{-1.15\baselineskip}}
\lstnewenvironment{curry}{\curryann \lstset{language=curry}}{}
\newcommand{\curryinline}[1]{\lstinline[language=curry,flexiblecolumns=false,mathescape=true,basewidth=0.55em]|#1|}

\newcommand{\code}[1]{\curryinline{#1}}

\lstnewenvironment{nolang}{\lstset{language=curry}}{}

\begin{document}

\title[A Monadic Implementation of Functional Logic Programs]
      {A Monadic Implementation of\\ Functional Logic Programs\footnote{%
This is a revised and extended version of \citep{HanusProttTeegen22},
invited as a rapid communication in TPLP.
The authors acknowledge the assistance of the conference program chairs
Beniamino Accattoli and Manuel Hermenegildo.}}

\lefttitle{Michael Hanus, Kai-Oliver Prott, Finn Teegen}

\begin{authgrp}
\author{Michael Hanus}
\affiliation{Institut f\"ur Informatik, Kiel University, Germany. \\
        \email{mh@informatik.uni-kiel.de}}
\author{Kai-Oliver Prott}
\affiliation{Institut f\"ur Informatik, Kiel University, Germany. \\
        \email{kpr@informatik.uni-kiel.de}}
\author{Finn Teegen}
\affiliation{Institut f\"ur Informatik, Kiel University, Germany. \\
        \email{fte@informatik.uni-kiel.de}}
\end{authgrp}

\maketitle

\begin{abstract}
Functional logic languages are a high-level approach
to programming by combining the most important declarative features.
They abstract from small-step operational details
so that programmers can concentrate on the logical aspects of an
application.
This is supported by appropriate evaluation strategies.
Demand-driven evaluation from functional programming is amalgamated
with non-determinism from logic programming
so that solutions or values are computed whenever they exist.
This frees the programmer from considering the influence
of an operational strategy on the success of a computation,
but it is a challenge to the language implementer.
A non-deterministic demand-driven strategy might duplicate
unevaluated choices of an expression, which could duplicate
the computational effort.
In recent implementations, this problem has been tackled
by adding a kind of memoization of non-deterministic choices
to the expression under evaluation.
Since this has been implemented in imperative target languages,
it was unclear whether this could also be supported in
a functional programming environment like Haskell.
This paper presents a solution to this challenge
by transforming functional logic programs into a monadic representation.
Although this transformation is not new,
we present an implementation of the monadic interface
which supports memoization in non-deterministic branches.
Additionally, we include more advanced features of functional logic languages,
namely functional patterns and encapsulated search, in our approach.
By optimizing our implementation for purely functional computations
with both a static and dynamic approach,
we are able to achieve a promising performance that outperforms
current compilers for Curry.

\medskip
\noindent
\textit{Under consideration in Theory and Practice of Logic Programming (TPLP)}

\end{abstract}


\section{Introduction}
\label{sec:introduction}

Declarative programming emphasizes the principle of expressing
properties of a given problem
in a high-level and execution-independent manner.
In functional programming languages,
equations specify the meaning of functions applied to given argument
patterns. These equations are used to reduce an initial expression to a value.
In logic programming languages, the meaning of predicates or relations
is specified by Horn formulas (implications).
The non-deterministic resolution principle \citep{Robinson65}
uses these formulas to compute solutions to a given query.

Functional logic languages \citep{AntoyHanus10CACM,Hanus13}
combine these programming paradigms in a single language environment.
In order to abstract from small-step operational details,
appropriate evaluation strategies are required.
Lazy or demand-driven strategies ensure
that iterated reduction steps w.r.t.\ given equations
compute a value if it exists \citep{HuetLevy91}.
Non-deterministic applications of program rules with overlapping
left-hand sides ensure that solutions are computed whenever they exist
\citep{Lloyd87}.
The combination of these techniques is called \emph{narrowing}
\citep{Slagle74,Reddy85}.
\emph{Needed narrowing} is a demand-driven variant which is optimal
w.r.t.\ the length of successful derivations and the
number of computed solutions \citep{AntoyEchahedHanus00JACM,Antoy97ALP}.

However, the combination of demand-driven and non-determin\-istic
evaluation steps might cause efficiency problems
in concrete implementations.
To sketch this problem, consider the following operations
written in Curry \citep{Hanus16Curry}\label{ex:not}\label{ex:aBool}\footnote{Since the syntax of Curry is close to
Haskell \citep{PeytonJones03Haskell} and we are going
to compile Curry programs into Haskell programs, we denote
the concrete language used in examples at the right margin.}:
\begin{curry}
not False = True       aBool = False ? True
not True  = False
\end{curry}
\code{not} is a standard function whereas \code{aBool}
is a \emph{non-determin\-istic operation} \citep{GonzalezEtAl99}
which uses Curry's archetypal \emph{choice} operation \ccode{?}
that returns one of its arguments.
Non-deterministic operations could have more than one value
for a given input,
e.g., \code{aBool} has values \code{False} and \code{True}.
They are an important concept of contemporary functional logic languages
(see \citep{AntoyHanus10CACM,GonzalezEtAl99} for more details).
Non-deterministic operations can be used in data structures or as arguments
to functions like any other operation, such as in \ccode{not$\;$aBool}.
Since the evaluation of any (sub)expression
might lead to a non-deterministic choice,
the occurrences of choices must be handled by the run-time system
of the language implementation.

Some implementations use \emph{backtracking}
to handle non-determinism, in particular,
implementations of functional logic languages that compile
into Prolog, like
PAKCS \citep{AntoyHanus00FROCOS,Hanus25PAKCS} or
TOY \citep{Lopez-FraguasSanchez-Hernandez99}.
Backtracking implements a choice by selecting one alternative
to proceed with the computation. If a computation terminates
(with success or failure), the state before the choice is restored, and
the next alternative is taken.
A well-known disadvantage of backtracking is its operational
incompleteness: if the first alternative does not terminate,
no result will be computed.
This problem can be avoided by keeping all alternatives in one computation
structure (e.g., a graph of expressions)
and using a fair strategy to explore this structure.

\emph{Pull-tabbing} is an approach to implement this idea.
It was first sketched in \citep{AlqaddoumiAntoyFischerReck10}
and formally explored in \citep{Antoy11ICLP}.
A pull-tab step is a local transformation
that moves a choice in a (demanded) argument of an operation
outside this operation. For instance,
\begin{nolang}
not (False ? True) $~\to~$ (not False) ? (not True)
\end{nolang}
is a pull-tab step.
Pull-tabbing is used in implementations targeting complete search
strategies, e.g., KiCS \citep{BrasselHuch07},
KiCS2 \citep{BrasselHanusPeemoellerReck11}, or
Sprite \citep{AntoyJost16}.
Iterated pull-tab steps move choices to the root of an expression.
If expressions containing choices are shared
(e.g., by applying rules with multiple occurrences of parameters
in their right-hand sides),
pull-tab steps might produce multiple copies of the same choice.
This could lead to unsoundness and duplication of computations.
The latter is a serious problem of pull-tabbing implementations
\citep{Hanus12ICLP}.
For instance, consider the additional operations\label{ex:xorSelf}
\begin{curry}
xor False x = x         xorSelf x = xor x x
xor True  x = not x
\end{curry}
From a logical point of view, \code{xor} applied to identical Boolean
values returns \code{False}.
Thus, \code{xorSelf} should always return \code{False}.
However, pull-tabbing transforms
the single choice occurring in the expression \code{xorSelf$\;$aBool}
into three choice occurrences.
\begin{nolang}
xorSelf aBool $~\to~$ xor aBool aBool $~\to~$ $\ldots$
              $~\to~$ (False ? True) ? (True ? False)
\end{nolang}
The value \code{True} (occurring twice) is incorrect.
Note that this is caused by the combination of lazy (demand-driven)
evaluation (which passes unevaluated expressions as arguments)
and non-determinism, which is evaluated on demand.
The problem of unsoundness can be fixed by
attaching identifiers to choices \citep{Antoy11ICLP}.
However, this does not fix the duplication of choices,
which can be detrimental to performance due to repeated evaluation.

The Curry compiler KiCS2 \citep{BrasselHanusPeemoellerReck11}
transforms Curry programs into Haskell programs and uses
pull-tabbing to handle non-determinism and to offer various search strategies.
Actually, KiCS2 attaches identifiers to choices and, thus,
suffers from the performance problem sketched above.
\citet{Hanus12ICLP} proposed an optimization which
eagerly evaluates demanded non-deterministic sub-expressions.
This optimization requires a demand analysis which is non-trivial
and often imprecise for complex data structures.

Recently, pull-tabbing has been improved
by adding a kind of memoization so that the re-evaluation
of shared non-deterministic choices is avoided.
This scheme, called \emph{memoized pull-tabbing} (MPT)
\citep{HanusTeegen21},
has been used in Curry implementations which transform source programs
into Julia programs \citep{HanusTeegen21}
or Go programs \citep{BoehmHanusTeegen21}.

Since MPT has been implemented only in imperative target languages,
it is still open to debate
whether the \enquote{MPT scheme can be combined with the
purely functional implementation approach of KiCS2} \citep{HanusTeegen21}.
A solution to this could be useful since KiCS2 supports maintainability
by its high-level target language and is also highly efficient
for purely functional computations
\citep{BrasselHanusPeemoellerReck11,BoehmHanusTeegen21}.

In this paper, we present a solution to this challenge
by transforming functional logic programs into a monadic representation.
Although such a transformation is not new \citep{Wadler90LFP,FischerKiselyovShan11},
we present an implementation of the monadic interface
which supports memoization of non-deterministic branches
and also advanced features of contemporary
functional logic languages,
such as functional patterns \citep{AntoyHanus05LOPSTR} and
encapsulation of non-determinism \citep{BrasselHanusHuch04JFLP}.

Our major contributions are:
\begin{enumerate}
  \item A basic monadic model for Curry that uses memoization
  to avoid repeated computations.
  \item Refinements of this basic model for efficiency improvements.
  \item Extensions to cover advanced features of functional logic languages.
\end{enumerate}
While we do not show the implementation of all extensions in detail
in this paper,
our implementation is the first Curry implementation that integrates
all major features of Curry with a promising performance.
A compiler that supports the full range of Curry features based on the
presented approach is available at {\small \url{https://github.com/Ziharrk/kmcc}}.

This paper is structured as follows.
The next section reviews some necessary details
about functional logic programming.
\cref{sec:lifting} presents the transformation
of functional logic programs into purely functional programs
parameterized by a monad to handle computational effects.
Some relevant implementations of pull-tabbing are sketched
in \cref{sec:history}, followed by the implementation of
our memoization monad in \cref{sec:memomonad}.
This monad is extended in \cref{sec:extensions} to cover
various features of contemporary functional logic languages,
like free (logic) variables and unification,
encapsulated search, and fair search strategies.
In \cref{sec:optimize} we present optimization techniques
to achieve a better performance for deterministic computations.
We evaluate our approach in \cref{sec:evaluation}
before we discuss related work in \cref{sec:related-work}
and conclude in \cref{sec:conclusions}.

\section{Functional Logic Programming}
\label{sec:flp}

We assume familiarity with basic ideas of logic and
functional programming and the language Haskell,
which we use for our implementation.
In the following, we review only concepts of
functional logic languages which are addressed
by the implementation presented later.
Concrete examples are shown in the multi-paradigm
declarative language Curry.\footnote{\url{https://www.curry-lang.org}}
More details can be found in surveys on
functional logic programming \citep{AntoyHanus10CACM,Hanus13}
and in the language report \citep{Hanus16Curry}.

Functional logic languages amalgamate distinguishing features
from functional programming
(demand-driven evaluation, strong typing with parametric
polymorphism, higher-order functions) and logic programming
(non-determinism, computing with partial information, constraints).
The language Curry has a Haskell-like syntax\footnote{%
Variables and function names usually
start with lowercase letters and the names of type and data constructors
start with an uppercase letter. The application of $f$
to $e$ is denoted by juxtaposition (``$f~e$'').}
\citep{PeytonJones03Haskell} but allows
\emph{free} (\emph{logic}) \emph{variables}
in conditions and right-hand sides of defining rules.
In contrast to Haskell and similarly to logic programming,
rule selection is non-deterministic,
i.e., if more than one rule is applicable, all applicable rules are tried.
The operational semantics is based on an optimal evaluation strategy
\citep{Antoy97ALP,AntoyEchahedHanus00JACM}---a conservative extension
of lazy functional programming and logic programming.

As an example, consider the ``classical'' functional logic definition
of the operation \code{last} to compute the last element of a list
(``\code{++}'' is the standard list concatenation operator,
\ccode{=:=} denotes unification, and free variables are introduced
by the keyword \code{free}):\label{example:last}
\begin{curry}
last xs | ys ++ [x] =:= xs = x
  where x,ys free
\end{curry}
This definition uses a conditional rule where the condition
is solved by evaluating \code{ys ++ [x]}
(which instantiates \code{ys} to some list)
and unifying the result with the input list \code{xs}.

As mentioned in \cref{sec:introduction},
\emph{non-deterministic operations} \citep{GonzalezEtAl99}
are an important feature of contemporary functional logic languages.
They are conceptually equivalent to free variables
(as shown in \citep{AntoyHanus06ICLP})
and can be nested like other functions.
This is due to the fact that non-deterministic operations
return (non-deterministically) individual values rather than sets of values.
Their declarative meaning can be specified
with a set-valued semantics \citep{GonzalezEtAl99}.
For instance, consider the following operation
that inserts an element at an unspecified position
into a list:\label{ex:insert}
\begin{curry}
insert :: a -> [a] -> [a]
insert x []     = [x]
insert x (y:ys) = (x : y : ys) ? (y : insert x ys)
\end{curry}
Hence, the expression \code{insert$\;$0$\;$[1,2]}
non-deterministically evaluates to one of the values
\code{[0,1,2]}, \code{[1,0,2]}, or \code{[1,2,0]}.
One can use this operation to easily define permutations:
\begin{curry}
perm :: [a] -> [a]
perm []     = []
perm (x:xs) = insert x (perm xs)
\end{curry}
Although \code{perm} is defined by non-overlapping rules,
the use of \code{insert} has the effect that
\code{perm$\;$[1,2,3,4]} non-deterministically evaluates to
all 24 permutations of the input list.

Compared to approaches where sets or lists of values
are passed between operations,
as in the \enquote{list of successes} approach
in purely functional programming \citep{Wadler85},
non-deterministic operations lead to simpler program structures.
Furthermore, they have operational advantages:
since expressions are evaluated on demand,
non-deter\-ministic operations as arguments result
in a demand-driven construction of the search space,
leading to considerably smaller search spaces
(see \citep{AntoyHanus10CACM,GonzalezEtAl99,Hanus24LOPSTR}
for more detailed discussions).

As mentioned in \cref{ex:xorSelf}, the occurrence of
non-deterministic operations as arguments
might yield incorrect results when such arguments are rewritten
to multiple occurrences.
For instance, if we evaluate the expression \code{xorSelf$\;$aBool} as in standard term rewriting
\citep{BaaderNipkow98}, i.e., by applying program rules from left to right,
there is the derivation (among others)\label{ex:xorSelfaBool}
\begin{curry}
xorSelf aBool  $\to~$  xor aBool aBool
               $\to~$  xor True aBool
               $\to~$  xor True False
               $\to~$  not False
               $\to~$  True
\end{curry}
The result \code{True} is incorrect, as discussed in \cref{ex:xorSelf}.
It is interesting to note that
this value cannot be obtained with a strict evaluation strategy
where arguments are evaluated prior to the function calls.
To avoid dependencies on the rewriting strategy
and exclude such incorrect results,
\citet{GonzalezEtAl99} proposed
the rewriting logic CRWL as a logical
(execution- and strategy-independent) foundation for declarative
programming with non-strict and non-deterministic operations.
CRWL specifies the
\emph{call-time choice} semantics \citep{Hussmann92}\label{ctc-semantics},
where values of the arguments of an operation are determined before the
operation is evaluated. This can be enforced in a lazy strategy
by sharing actual arguments.
For instance, the expression above can be lazily evaluated
provided that all occurrences of \code{aBool}
are shared so that all of them reduce either
to \code{False} or to \code{True} consistently.

Although sharing is also used in implementations of non-strict functional
languages, like Haskell,
in order to support optimal evaluation \citep{HuetLevy91},
we cannot directly map functional logic programs into Haskell programs
due to the non-deterministic features.
Thus, a correct mapping requires modeling non-determinism \emph{and}
sharing of non-deterministic expressions.
For this purpose, we will use a monadic representation of programs.

In addition to these base features, functional logic languages
have more features which are useful for application programming.
Apart from standard features like
modules or monadic I/O \citep{Wadler97},
\emph{encapsulation} \citep{BrasselHanusHuch04JFLP} is useful
to collect all results of a non-deterministic subcomputation
in some data structure,
and \emph{functional patterns} \citep{AntoyHanus05LOPSTR}
are useful to non-deterministically select sub-expressions
at arbitrary positions.
In the following, we first discuss a scheme to implement
demand-driven non-determinism in Haskell.
Later, we extend our scheme to more advanced features
in order to obtain a full-fledged implementation of Curry in Haskell.

\section{Monadic Transformation}
\label{sec:lifting}

In this section we present a basic implementation of
a kernel of Curry in Haskell.
When mapping a functional logic program into Haskell,
one has to model non-deterministic computations
in a functional manner.
A well-known method to represent non-deterministic results
in a functional language is the ``list of successes'' technique
\citep{Wadler85}: a non-deterministic operation is mapped
into a function which returns a list of values.
Instead of lists, one can also use other container structures, e.g., trees.
In order to abstract from the data structure to collect values,
we parameterize the target program by a monad.
A monad \code{m} is a type constructor with two operations
\begin{haskell}
return :: a -> m a
(>>=) :: m a -> (a -> m b) -> m b
\end{haskell}
To model failures and non-deterministic choices,
the more specific monadic structure \code{MonadPlus} is appropriate
since it offers two additional operations
\begin{haskell}
mzero :: m a
mplus :: m a -> m a -> m a
\end{haskell}
\code{mzero} represents a failing computation and \code{mplus}
a choice between two computations.
For instance, the non-deterministic operation \code{aBool}
defined in \cref{ex:aBool} can be mapped into%
\footnote{In order to distinguish Curry entities
from their translations into Haskell, we decorate the latter
with the suffix \ccode{C}.}
\begin{haskell}
aBoolC :: MonadPlus m => m BoolC
aBoolC = return FalseC `mplus` return TrueC
\end{haskell}
A simple instance of \code{MonadPlus} is the list monad
\citep{Wadler85} where \code{return} creates a singleton list,
\code{mzero} an empty list, and \code{mplus} concatenates the
argument lists. Then the monadic representation of \code{aBool}
returns the list of its two values:
\begin{haskell}
> aBoolC :: [BoolC]
[FalseC,TrueC]
\end{haskell}
The monadic bind operator \code{(>>=)} is used to pass
non-deterministic values of arguments.
For instance, consider the Curry expression \code{not$\;$aBool}.
To evaluate it, \code{not} must be applied to all values of \code{aBool}.
Thus, the monadic version of \code{not}
takes the monadic representation of its argument and returns
a monadic value:
\begin{haskell}
notC :: MonadPlus m => m BoolC -> m BoolC
notC x = x >>= \x' -> case x' of
                        FalseC -> return TrueC
                        TrueC  -> return FalseC
\end{haskell}
Since the bind operator of the list monad applies the second argument
to all elements of its first argument and concatenates all result values,
we can nest these operations:
\begin{haskell}
> notC (notC aBoolC) :: [BoolC]
[FalseC,TrueC]
\end{haskell}
Due to the monadic abstraction, one can also use other monad instances,
e.g., search trees, or add more effects to the monadic computation.
As shown later, this is the key to our implementation of advanced
functional logic programming features.

Because of these considerations,
we transform normal Curry code
to functional code parameterized by a monad
where all transformed operations get and return monadic values, which are non-deterministic computations in our case.
Such a transformation for call-by-name and call-by-value languages
has been presented by \citet{Wadler90LFP} and is also called \emph{monadic lifting}.

As sketched in \cref{sec:flp},
Curry uses a lazy call-by-need evaluation strategy.
However, Haskell's sharing is not sufficient to obtain call-by-need
for the monadic representation.
Since the bind operator triggers the evaluation of a monadic value,
multiple occurrences of the same expression might be independently evaluated.
For instance, the sub-expression \code{aBool} of the expression
\code{xorSelf$\;$aBool} (see \cref{ex:xorSelf})
has two occurrences in the further evaluation of this expression
which are independently evaluated.
In order to conform to the call-time choice semantics,
the non-deterministic values need to be shared.
While we could evaluate \code{aBool} before passing it to \code{xor} in \code{xorSelf},
this would only be correct because we know \code{xor} to be strict.
In general, such a call-by-value implementation is incorrect for a lazy language like Curry.
Thus, we explicitly model sharing using an approach adapted from
both \citet{FischerKiselyovShan11} and \citet{Petricek12}.
For this, we need to introduce an operator\label{def:share-fischer}
\begin{haskell}
share :: Monad m => m a$\;$-> m (m a)
\end{haskell}
By passing a \enquote{to-be-shared} monadic expression to \code{share}
and extracting the result using \code{(>>=)},
we obtain a new monadic expression that respects
our call-by-need evaluation strategy.
Consider the following translation of \code{xorSelf}
which uses its argument twice.
\begin{haskell}
xorSelfC x = share x >>= \x' -> xorC x' x'
\end{haskell}
On the first evaluation of the new variable \code{x'} inside \code{xorC},
the evaluation of the original argument \code{x} to \code{share}
is triggered and the computed result is memoized.
On any subsequent evaluation of \code{x'},
the memoized result of \code{x} is used without evaluating \code{x} again.

Now we will present the transformation of Curry code to an explicit
monadic variant.  For the rest of this paper, we will use \code{Curry}
to denote our monadic effect type and the following type for transformed
functions to increase readability.
\begin{haskell}
newtype a :-> b = Func (Curry a -> Curry b)
\end{haskell}

\subsubsection*{Types}
On the type level, the monadic transformation replaces type constructors with
their effectful counterparts and wraps the result and argument of each
function type in \code{Curry}.
However, any type quantifiers and type constraints (which might be absent)
still remain at the beginning
of the type signature and we also wrap the outer type of a function.
\cref{fig:lifting-types} presents the transformation of types, including the
renaming of type constructors by the operation ``\rename''
(e.g., add the suffix \ccode{C}).
For instance, the type signature
\begin{curry}
not :: Bool -> Bool
\end{curry}
is transformed into
\begin{haskell}
notC :: Curry (BoolC :-> BoolC)
\end{haskell}
The transformed type signature differs from the one we gave previously
for \code{notC}. The key difference is that we wrap the monadic type
constructor \code{Curry} around the whole type. The latter is
required because a function could introduce non-determinism before
being applied to an argument.  To see this, consider the following
artificial function that is non-deterministically defined as either
identity or negation on Boolean values.
\begin{curry}
idOrNot :: Bool -> Bool
idOrNot = id ? not
\end{curry}
The transformed type signature has the form
\begin{haskell}
idOrNotC :: Curry (BoolC :-> BoolC)
\end{haskell}
For this function, having the monadic type constructor \code{Curry}
at the outer level is necessary.
Since we want to decide how to transform the type of a function based
solely on its type and not on its implementation, we
treat all functions as potentially introducing non-determinism.
Note that inlining and rewrite optimizations can get rid of some of the
overhead introduced here.
Such optimizations are possible with the Glasgow Haskell Compiler
that we target.

\begin{figure*}
\begin{minipage}[c]{\textwidth}
  \begin{align*}
  \mcode{$\llbracket \forall \alpha_1 \ldots \alpha_n$. $\phi \Rightarrow \tau \rrbracket^t$}
    &\coloneqq \mcode{$\forall \alpha_1 \ldots \alpha_n$. $\llbracket \phi \rrbracket^t \Rightarrow$ Curry ($\llbracket \tau \rrbracket^i$)} \tag{Polymorphic type}\\
  \mcode{$\llbracket \langle \kappa_1  \;\tau_1$, $\ldots$, $\kappa_n \;\tau_n \rangle \rrbracket^t$}
    &\coloneqq \mcode{$\langle \rename$($\kappa_1$)$\llbracket \tau_1 \rrbracket^i$, $\ldots$, \rename($\kappa_n$)$\llbracket \tau_n \rrbracket^i \rangle$} \tag{Context}\\
  \end{align*}
  \vspace{-1.75em}
  \begin{align*}
  \mcode{$\llbracket \tau_1 \rightarrow \tau_2 \rrbracket^i$}
    &\coloneqq \mcode{$\llbracket \tau_1 \rrbracket^i \funCurry \llbracket \tau_2 \rrbracket^i$} \tag{Function type}\\
  \mcode{$\llbracket \tau_1 \;\tau_2 \rrbracket^i$}
    &\coloneqq \mcode{$\llbracket \tau_1 \rrbracket^i$ $\llbracket \tau_2 \rrbracket^i$} \tag{Type application}\\
  \mcode{$\llbracket \chi \rrbracket^i$}
    &\coloneqq \mcode{\rename($\chi$)} \tag{Type constructor}\\
  \mcode{$\llbracket \alpha \rrbracket^i$}
    &\coloneqq \mcode{$\alpha$} \tag{Type variable}
  \end{align*}
\end{minipage}
\caption{Type transformation $\llbracket \circ \rrbracket^t$}
\label{fig:lifting-types}
\end{figure*}

\begin{figure*}
\begin{minipage}[c]{\textwidth}
  \begin{align*}
  \mcode{$\llbracket$\textbf{data} $D$ $\alpha_1 \ldots \alpha_n$ = $C_1$ | $\ldots$ | $C_n$ $\rrbracket^d$}
    &\coloneqq \mcode{\textbf{data} \rename($D$) $\alpha_1 \ldots \alpha_n$ = $\llbracket C_1 \rrbracket^c$ | $\ldots$ | $\llbracket C_n \rrbracket^c$}\tag{Data type}\\
  \mcode{$\llbracket C$ $\tau_1 \ldots \tau_n \rrbracket^c$}
    &\coloneqq \mcode{\rename($C$) $\llbracket \tau_1 \rrbracket^t \ldots \llbracket \tau_n \rrbracket^t$}\tag{Constructor}
  \end{align*}
\end{minipage}
\caption{Data type transformation $\llbracket \circ \rrbracket^d$}
\label{fig:lifting-data}
\end{figure*}

\subsubsection*{Data Type Declarations}
\label{sec:data-type-declarations}

Due to the non-strictness of Curry, components of data types
may also contain non-determinism.
As discussed in \citep{FischerKiselyovShan11} in detail,
this can be achieved by transforming also
the arguments of data constructors into monadic values.
Thus, we need to modify data type definitions
(except for primitive data types since they have no components).
Because we rename the data types during the transformation,
we have to update any type constructor in a type to use the name
of its effectful counterpart.
We take a look at the transformation of data types next.
To support data constructors having unevaluated or non-deterministic
components, we transform every constructor.
As the (partial or full) application of a constructor can never
be non-deterministic by itself, we neither have to transform the result type
of the constructor nor wrap the function arrow.
This allows us to transform only the parameters of constructors,
because they are the only potential sources of
non-deterministic effects in a data type.
\cref{fig:lifting-data} shows the rules for the transformation of data types.
The following code shows an example for this transformation.
To improve readability, we use regular algebraic data type syntax
for the list type and its constructors instead of
the special list syntax of Curry and Haskell.
The Curry data type
\begin{curry}
data List  a = Nil
             | Cons a (List a)
\end{curry}
is transformed into
\begin{haskell}
data ListC a = NilC
             | ConsC (Curry a) (Curry (ListC a))
\end{haskell}

\noindent
Because the transformation of a constructor differs from the transformation of a function,
we later have to treat constructors and functions in expressions differently.
We rename constructors and type constructors in our implementation
so that we can more easily identify a transformed (type) constructor.

\subsubsection*{Functions}
\label{sec:lifting-functions}
The type of a transformed function serves as a guide for the transformation of functions and expressions.
We can derive three rules for our transformation of expressions from our transformation of types:
\begin{enumerate}
  \item Each function arrow is wrapped in a \code{Curry} type.
  Therefore, function definitions are replaced by constants that use
  a sequence of lambda expressions to introduce the arguments
  of the original function.
  All lambda expressions are wrapped in a \code{return} because function arrows are wrapped in a \code{Curry} type.
  \item We have to extract a value from the monad using \code{(>>=)}
  before we can pattern match on it.
  As a consequence, we cannot pattern match directly on the argument
  of a lambda or use nested pattern matching, as arguments of a
  lambda and nested values are effectful and have to be extracted
  using \code{(>>=)} again.
  \item Before applying a function to an argument, we first have to extract the function from the monad using \code{(>>=)}.
  As each function arrow is wrapped separately, we need this extraction for every parameter applied to the original function.
\end{enumerate}
Implementing this kind of transformation in one pass over
a concrete Curry program is challenging.
Thus, we assume that a source program is simplified first.
For the remainder of this subsection, we assume that our program
is already desugared into a flat form:\footnote{The flat form of
programs is also used for the semantics \citep{AlbertHanusHuchOliverVidal05},
analysis \citep{HanusSkrlac14}, and
implementation \citep{AntoyHanusJostLibby20} of
functional logic programs.}
a program is a set of top-level
functions in the form of lambda abstractions (nested definitions
are removed by lambda lifting \citep{Johnsson85}),
pattern matching is represented with non-nested patterns in case expressions,
all cases branches are complete (i.e., branches on missing constructors
are completed with \code{failed}, the predefined always failing operation),
and non-determinism (e.g., overlapping rules)
is expressed with the choice operator \code{(?)}.
As shown by \citet{Antoy01PPDP},
any functional logic program or constructor-based conditional
term rewriting system can be transformed into this flat form.
For instance, the operation \code{insert} defined in \cref{ex:insert}
is desugared into the following flat form.
\begin{curry}
insert = \x -> \xs -> case xs of
  []   -> [x]
  y:ys -> (x : y : ys) ? (y : insert x ys)
\end{curry}
\cref{fig:transformation-expressions} presents the rules of our
monadic transformation on programs in flat form.
We explain the rules for abstractions and applications in more detail.
Most of the other rules are straightforward.
Since our transformation only works for correctly typed programs,
we assume that the input code to our transformation has been checked
for type errors.

\begin{figure*}
\begin{minipage}[c]{\textwidth}
  \begin{align*}
  \mcode{$\llbracket x \rrbracket^e$}
    &\coloneqq x &\tag{Variable}\\
  \mcode{$\llbracket f \rrbracket^e$}
    &\coloneqq \rename(f) &\tag{Defined Function}\\
  \mcode{$\llbracket e_1 \; e_2 \rrbracket^e$}
    &\coloneqq \mcode{$alias(\llbracket e_2 \rrbracket^e,\, y,\; $apply$\, \llbracket e_1 \rrbracket^e$\;$y)$} \tag{Application}\\
  \mcode{$\llbracket \lambda x \rightarrow e \rrbracket^e$}
    &\coloneqq \mcode{return (Func ($\lambda x \rightarrow \llbracket e \rrbracket^e$))} &\tag{Abstraction}\\
  \mcode{$\llbracket C \rrbracket^e$}
    &\coloneqq \mcode{return (Func ($\lambda y_1 \rightarrow \ldots$ return (Func ($\lambda y_n \rightarrow$} & \; \\
    & \ \ \ \ \ \; \mcode{return (\rename($C$) $y_1 \ldots y_n)))\ldots))$} &\tag{Constructor}\\
  \mcode{$\llbracket$\textbf{let} $x$ = $e_1$ \textbf{in} $e_2$$\rrbracket^e$}
    &\coloneqq \mcode{\textbf{let} $y$ = $\llbracket$$e_1$$\rrbracket^e$ \textbf{in} $alias(y,x,\llbracket e_2 \rrbracket^e)$} \tag{Let Expression}\\
  \mcode{$\llbracket$(?)$\rrbracket^e$}
    &\coloneqq \mcode{return (Func ($\lambda x \rightarrow$ return (Func ($\lambda y \rightarrow$ $x$ `mplus` $y$))))} &\tag{Choice operator}\\
    \mcode{$\llbracket$failed$\rrbracket^e$}
    &\coloneqq \mcode{mzero} &\tag{Failed computation}\\
  \mcode{$\llbracket$\textbf{case} $e$ \textbf{of} $\{ br_1; \ldots; br_n \}\rrbracket^e$}
    &\coloneqq \mcode{$\llbracket e \rrbracket^e$ $\bind$ $\lambda y \rightarrow$}\ \mcode{\textbf{case} $y$ \textbf{of} $\{ \llbracket br_1 \rrbracket^b; \ldots; \llbracket br_n \rrbracket^b \}$}&\tag{Case Expression}\\
  \mcode{$\llbracket C~x_1 \ldots x_n \rightarrow e \rrbracket^b$}
    &\coloneqq \mcode{\rename($C$) $x_1 \ldots x_n \rightarrow$}\ \mcode{$\llbracket e \rrbracket^e$} &\tag{Case Branch}\\[1em]
  \mcode{$alias(e_1, x, e_2)$}
    &\coloneqq \mcode{share $e_1 \; \bind$ $\lambda x \rightarrow e_2$} &\tag{Aliasing}
  \end{align*}
\end{minipage}
\caption{Expression transformation $\llbracket \circ \rrbracket^e$ (where $f$, $y$, $y_1,\ldots,y_n$ are fresh variables)}
\label{fig:transformation-expressions}
\end{figure*}

\paragraph{Applications}
For better readability, the transformation of a function application
uses the auxiliary function
\begin{haskell}
apply mf mx = mf >>= \(Func f) -> f mx
\end{haskell}
which extracts the \enquote{real} function from the monad and applies
it to its monadic argument.
Thus, a transformed function application applies the transformed function
to the transformed and shared argument.
An application of a function to more than one argument is represented as multiple nested applications.
Similar to \citep{Petricek12}, we use \code{share} in the application of a function
to an argument, instead of whenever a variable is brought into scope like in \citep{FischerKiselyovShan11}.
This also allows us to get rid of the \enquote{deep sharing}
that \citet{FischerKiselyovShan11} required.
In contrast to \citep{Petricek12},
we also consider data types, as well as let and case expressions.
Since constructor applications are shared via the application rule,
the constructor arguments in a pattern of a case expression are already shared.
Let expressions, however, introduce new variables that should be shared across
different application sites in the body of the let expression.
Therefore, we need to explicitly use \code{share} in their translation.

\paragraph{Lambda abstractions}
A lambda abstraction is translated by wrapping it in a \code{return}
as well as the transformed function constructor \code{Func} and translating
the inner expression.
We do not need to apply \code{share} to the argument of the lambda,
since any argument supplied to a function is already shared
by the rule for applications.

\paragraph{Transformation examples}
As an example, consider the operation \code{not} defined in \cref{ex:not}.
The flat form of \code{not} is
\begin{curry}
not = \x -> case x of False -> True
                      True  -> False
\end{curry}
This is transformed into
\begin{haskell}
notC = return (Func (\arg -> arg >>= \x ->
                       case x of FalseC -> return TrueC
                                 TrueC  -> return FalseC ))
\end{haskell}
As a further example with more complex pattern matching and applications,
we show the transformation of the \code{perm} operation
from \cref{sec:flp} where we reduced some $\eta$-redexes for readability.
\begin{haskell}
permC :: Curry (ListC a :-> ListC a)
permC = return (Func (\arg -> arg >>= \xs ->
           case xs of
             NilC       -> return NilC
             ConsC y ys -> share (share ys >>= apply permC) >>=
                             apply (share y >>= apply insertC) ))
\end{haskell}
Even though the code is now significantly more complex,
the user will not have to read or write such code
since the transformation can be fully automated.
Another thing to note is that the applications of \code{share} on \code{y}
and \code{ys} are superfluous in this code snippet.
The reason for this and how to avoid these superfluous applications will be discussed in \cref{sec:performance}.

\section{A History of Monadic Pull-Tabbing}
\label{sec:history}

This section discusses some existing monadic implementations
of pull-tabbing and their performance deficiencies.
Our own implementation builds on some of these ideas but fixes
the problems they have.

\subsection{Tree-based Non-determinism with Fingerprinting}

As already discussed in \cref{sec:lifting},
the monadic transformation of functional logic programs
supports different implementations of non-deterministic computations by providing different instances of \code{MonadPlus}.
The list instance computes a list of all non-determin\-istic values
and corresponds to backtracking search used in Prolog.

In order to support other search strategies,
one can collect non-deterministic values in tree structures
rather than lists.
For this, we use a standard binary tree representation as seen below.
\label{def:data-tree}
\begin{haskell}
data Tree a = Empty
            | Leaf a
            | Node (Tree a) (Tree a)
\end{haskell}
Thus, \code{Leaf} represents a single value,
\code{Empty} no value, i.e., a failure, and
\code{Node} a non-deterministic choice between trees of values.
With this intuition in mind, it is straightforward to define
the \code{Monad} and \code{MonadPlus} instances as follows.

\begin{haskell}
instance Monad Tree where
  return = Leaf $\listline$
  Empty      >>= _ = Empty
  Leaf x     >>= f = f x
  Node t1 t2 >>= f = Node (t1 >>= f) (t2 >>= f)

instance MonadPlus Tree where
  mzero = Empty$\listline$
  mplus = Node
\end{haskell}
Using the monadic transformation of functional logic programs
with this tree monad, each operation computes a tree of
non-deterministic values.
By applying different tree traversals,
one can easily implement different search strategies,
like depth- or breadth-first search.
In practice, this is used in the Curry implementation
KiCS2 \citep{BrasselHanusPeemoellerReck11}
which has options to select various search strategies
(e.g., depth-/breadth-first, iterative deepening, parallel)
\citep{HanusPeemoellerReck12ATPS}.%
\footnote{The actual implementation of KiCS2
is not based on a monadic representation of Curry programs
but uses a direct encoding of search trees in translated operations.}

Using just a tree for the monadic effect is not sufficient to implement
a call-time choice semantics.
The tree-based monadic non-determinism yields unintended answers
(as with standard term rewriting, see \cref{ex:xorSelfaBool}):
\begin{haskell}
> xorSelfC aBoolC :: Tree BoolC
Node (Node (Leaf FalseC) (Leaf TrueC))
     (Node (Leaf TrueC) (Leaf FalseC))
\end{haskell}
Here we can observe the problem discussed in \cref{sec:introduction},
namely the duplication of choices,
which potentially leads to unsoundness w.r.t.\ call-time choice
and duplicated computations.
The former can be avoided by attaching identifiers to choices.
When a choice is created by an occurrence of the operation \ccode{?},
it is decorated with a fresh \emph{choice identifier}.
In a pull-tab step, i.e., when a non-deterministic demanded argument
causes a choice of results, the choice identifier is passed from
the argument to the result.
In our example, all three \code{Node} constructors in the result tree
would be decorated with the same choice identifier.
If the tree traversal that extracts result values from a search tree
always makes \emph{consistent} choices, i.e., selects the same
(left/right) branch for identically decorated choices,
then the computed values are the intended ones \citep{Antoy11ICLP}.
Therefore, implementations of functional logic languages
which use pull-tabbing to support flexible and operationally complete
search strategies \citep{BrasselHanusPeemoellerReck11,AntoyJost16}
often require choice identifiers in computations.
A \emph{computation branch}, which evaluates an expression with
some decision for non-deterministic choices, contains
a \emph{fingerprint}, a partial mapping from choice
identifiers to left/right decisions, and computes
values for this fingerprint in a call-by-need manner.
This is done in KiCS2 where the search tree traversal
is parameterized by fingerprints \citep{BrasselHanusPeemoellerReck11}.

Unfortunately, this method to implement non-determinism
could cause a serious efficiency problem.
Since pull-tabbing moves every choice occurring in arguments
to the root of an expression, choices occurring in shared
subexpressions are multiplied (before they are removed by fingerprinting).
For instance, KiCS2 is the most efficient Curry implementation
on purely functional programs \citep{BrasselHanusPeemoellerReck11}
but it might be much slower than other Curry implementations
for particular uses of non-deterministic operations \citep{Hanus12ICLP}.

\subsection{Explicit Sharing of Computations}
A solution to pull-tabbing without fingerprinting is proposed
by \citet{FischerKiselyovShan11}.
Their implementation not only solves the problem of unsoundness
but also aims to avoid the duplication of computations.
They use the \code{share} operation
we mentioned in \cref{sec:lifting} to save the result
of a potentially non-deterministic computation on a heap
local to the current computation branch.
While their approach makes a key step in the right direction,
it suffers from a flaw that decreases performance in real-world applications.
The fact that results are only stored and looked up locally
in the current computation branch implies that results
cannot be shared across non-deterministic branches.
Consider the following code snippet.\label{ex:primes}
\begin{curry}
primes :: [Int]
primes = $<\textrm{\!deterministic definition of an infinite list of primes\!}>\listline$
sharingAcrossND :: Int
sharingAcrossND = let prime800 = primes !! 799
                  in prime800 ? prime800
\end{curry}
The value of \code{prime800} will be computed for the first time
when one argument of the choice operator \code{(?)} is evaluated.
At that point, the value of \code{prime800} is stored on the heap
only in that computation branch.
Thus, we do not have the computed value available on the heap
of the alternative branch.
When evaluating the other argument of the choice,
the value of \code{prime800} will be required again.
Consequently, we need to evaluate \code{prime800} a second time
even though it will yield the same result.
In conclusion, deterministic values are not shared
across non-deterministic branches
in this setting, as already noticed by \citet{FischerKiselyovShan11}.
It should be noted that the same problem occurs in a backtracking-based
implementation of non-determinism: a computation performed
in one non-deterministic branch is undone before an alternative
non-deterministic branch is selected.

\section{A Non-determinism Monad with Memoization}
\label{sec:memomonad}

In this section we introduce the kernel of our implementation
by developing an implementation of a basic memoized
non-determinism monad that aims to fix the problems
discussed in the previous section.
It is based on a solution to pull-tabbing without fingerprinting
that enables sharing across non-determinism,
proposed in \citep{HanusTeegen21},
but here the results of non-deterministic computations are memoized
in a different way so that it fits into the functional monadic transformation.

An implementation based on the monadic transformation
presented in \cref{sec:lifting} has to implement
the functional interface on which this transformation is based,
i.e., an appropriate monad instance and
an implementation of the operation \code{share}.
This is quite similar to the approach of \citet{FischerKiselyovShan11}
since they proposed the same interface.
While they give a purely functional implementation of this interface
for lazy non-determinism,
their approach suffers from the mentioned drawbacks.
In order to avoid these, our implementation uses a mutable (global) state
to implement memoization. Thus, we implement the same interface
(monad and \code{share}) but hide behind the interface
an impure implementation for the sake of efficiency.

In standard implementations of non-strict functional languages,
the node of a computation graph representing an operation
is updated in-place with the computed result in order to share it.
This is not possible in a functional logic language with an implementation based on pull-tabbing,
since an operation might have more than one result.
To overcome this problem, computation branches are uniquely identified
by \emph{branch identifiers}.
Instead of updating a node in-place,
potentially non-deterministic operation nodes
contain a (partial) map $\tr$, called \emph{task result map},
from branch identifiers to results.
When a branch $i$ has to evaluate some node $n$ containing
an operation with map $n.\tr$,
it first checks whether $n.\tr(i)$ is defined.
If not, node $n$ is evaluated and $n.\tr(i)$ is set to the computed result
before returning $n.\tr(i)$.
This evaluation scheme, called
\emph{memoized pull-tabbing} (MPT) \citep{HanusTeegen21}
and formally described in \citep{BoehmHanusTeegen21},
avoids the aforementioned problem of pull-tabbing.
Its efficiency can compete with backtracking-based implementations
 \citep{HanusTeegen21}.
Furthermore, it is a good basis for operationally complete
implementations of functional logic languages.
Existing implementations \citep{BoehmHanusTeegen21,HanusTeegen21}
use imperative target languages.
In the following, we discuss a high-level Haskell-based implementation
that exploits the monadic transformation introduced so far.

\subsection{Memoization Monad}

To extend the tree-based monadic non-determinism implementation
with memoization for call-by-need evaluation and call-time-choice,
we need storages for memoized results and identifiers to uniquely label every branch of
a non-deterministic computation.
For the latter, we assume an abstract type of identifiers
that support a \code{nextID} operation to generate new identifiers.
\begin{haskell}
type ID
nextID :: ID -> ID
\end{haskell}
Based on this abstract type, we define a state to store
the current branch identifier and any \enquote{parent} identifier
that occurred higher up in our tree of non-deterministic choices.
The parents are necessary because a deterministic result that
was memoized for a certain branch identifier is also valid in any child branch.
Finally, we need an \code{IORef} to be able to generate \code{ID}s
that are unique across all branches.
The \code{IORef} itself will not change during computations
but its content will.
For example, the operation \code{freshID} below
updates the content of this \code{IORef} to generate a new identifier.
\begin{haskell}
data MemoState = MemoState {
    branchID  :: ID,
    parentIDs :: Set ID,
    idSupply  :: IORef ID
  }$\listline$
PRAGMAFRESHID
freshID :: MonadState MemoState m => m ID
freshID = do
  MemoState _ _ idSupply <- get
  let val = unsafePerformIO <>
              atomicModifyIORef idSupply (\i -> (nextID i, i))
  return val
\end{haskell}
Note that we need a \code{NOINLINE} pragma on \code{freshID} so that
optimizations will not interfere with our usage of \code{unsafePerformIO}.
Instead of using unsafe features directly,
we could also use the \code{UniqSupply} type that is used inside
the Glasgow Haskell Compiler (GHC) itself,
but that one uses unsafe features under the hood as well.

Below we define the type of our basic Curry monad.
It is just a state monad on top of our tree structure.
\begin{haskell}
newtype Curry a = Curry {
    unCurry :: StateT MemoState Tree a
  } deriving (Functor, Applicative, Monad)
\end{haskell}
The \code{Monad} instance can be derived from the underlying
\code{StateT} implementation using generalized \code{newtype} deriving, but
we explicitly need to give the \code{MonadPlus} instance so that we
can manipulate our branch identifiers correctly.
\begin{haskell}
instance MonadPlus Curry where
  mzero = Curry (lift Empty)

  Curry m1 `mplus` Curry m2 = Curry <> do
    MemoState branchID parentIDs idSupply <- get
    i1 <- freshID
    i2 <- freshID
    let newPs = Set.insert branchID parentIDs
        leftState  = MemoState i1 newPs idSupply
        rightState = MemoState i2 newPs idSupply
    (put leftState >> m1) `mplus` (put rightState >> m2)
\end{haskell}

\subsection{Memoizing Non-determinism}
Now we define the central function \code{share}
that enables lazy evaluation via memoization.
\begin{haskell}
share :: Curry a -> Curry (Curry a)
\end{haskell}
Compared to the type described by \citet{FischerKiselyovShan11},
we have removed a type class constraint.
As mentioned before, \citet{FischerKiselyovShan11} need this additional type class
to ensure that expressions nested deep inside a data structure
are shared and memoized as well (referred to as \enquote{deep sharing}).
However, since we use \code{share} at every (constructor) application site,
we already get this deep sharing for free.
The function \code{share} does the actual memoization work.
When called with an argument computation,
it first creates a new task result map to memoize possible
results of this computation.
Then it returns a new computation which checks and updates
this map accordingly.
We present the implementation of \code{share} step by step.
\begin{haskell}
share :: Curry a -> Curry (Curry a)
share (Curry ma) = Curry <> do
  let trRef = unsafePerformIO (newIORef emptyHeap)
  MemoState b1 _ _ <- get
  return <> $\ldots$ -- see below
\end{haskell}
To store memoized results,
we use the following data type \code{Heap},
a mapping from \code{ID}s to values,
with both an \code{insertHeap} and \code{lookupHeap} operation.
The implementation of the heap is not relevant,
as it can be any kind of key-value store.
\begin{haskell}
type Heap a
emptyHeap  :: Heap a
insertHeap :: ID -> a -> Heap a -> Heap a
lookupHeap :: ID -> Heap a -> Maybe a
\end{haskell}
We initialize the heap by using \code{unsafePerformIO}
together with \code{newIORef$\;$emptyHeap} in the function \code{share}
to create a kind of memory cell where we can manipulate our results.%
\footnote{We tried getting rid of \fcode{IO}
in our implementation but we conjecture that this is not possible for our goal.
Note that the usage of \fcode{unsafePerformIO} is captured
in a safe interface here and is, thus, just ugly but still safe.}
In the next step, the current branch ID (\code{b1}) is retrieved from the state
so that we can use the correct ID to insert into the task result map.
This is everything that we need to do to prepare a shared computation.
The actual execution of our computation \code{ma} happens in the
returned computation which starts with the \code{return} in the last line.
We continue at that \code{return}.
\begin{haskell}
  return <> do
    MemoState b2 p2 idSupply2 <- get
    case lookupTaskResult trRef b2 p2 of
      Nothing  -> $\ldots$ -- see below
      Just res -> $\ldots$ -- see below
\end{haskell}
Here, we retrieve the current (possibly different) state and check
if we already have a result saved in the task result map
that is valid in the current or a parent branch.
Checking the map for a memoized result is done using \code{lookupTaskResult}
which is defined as follows.
\begin{haskell}
lookupTaskResult :: IORef (Heap a) -> ID -> Set ID -> Maybe a
lookupTaskResult trRef i p =
    msum (map (\j -> lookupHeap j h) (i : Set.toList p))
  where h = unsafePerformIO (readIORef trRef)
\end{haskell}
The current map is read using \code{unsafePerformIO} and,
for the current and each parent branch ID, we check
if there is a valid result memoized in this map.
Combining all these \code{Maybe$\;$a} results using the monoidal \code{msum},
we obtain the first result if it exists, or \code{Nothing} otherwise.
Note that there can only ever be at most one valid result in the map.

Now we continue with the \code{Nothing} case of \code{share}
where no valid memoized result exists.
\begin{haskell}
      Nothing -> do
        a <- ma
        MemoState b3 _ _ <- get
        let wasND = b2 /= b3
        let whereTo = if wasND then b3 else b1
        let addTR h = insertHeap whereTo a h
        unsafePerformIO (modifyIORef trRef addTR)
          `seq` return a
\end{haskell}
Here, we have to execute the computation \code{ma} to obtain a result
that we can memoize.
Immediately afterwards, we get the current branch ID (\code{b3})
so that we can check if the computation \code{ma} contained non-determinism.
If the branch ID changed during its execution (i.e., \code{b2$\;$/=$\;$b3}),
the computation was indeed non-deterministic and, thus,
the new result is valid only for the local branch identifier \code{b3}.
In the case that the branch ID did not change,
\code{ma} was deterministic and, thus, the result is globally valid
for the branch identifier \code{b1} that was active
on the outer monadic layer of \code{share}.
With this information at hand, we construct a new task result map and
update the \code{IORef} using \code{unsafePerformIO}.
We have to force the computation of the result of \code{unsafePerformIO}
using \code{seq} to ensure that it actually happens
\emph{before} the result of the computation \code{ma} is returned.

If we actually have a result memoized,
the \code{Just} branch after our lookup simply returns the result.
\begin{haskell}
      Just res -> return res
\end{haskell}

\subsection{Nested Sharing of Non-deterministic Values}
The implementation given above is correct for programs
where non-deterministic values are not shared twice or more.
However, there are some programs where such a multiple sharing happens.
To illustrate the problem, consider the following Curry program.
\begin{curry}
notIf :: Bool -> Bool
notIf x = let nx = not x
          in if x then nx else nx
\end{curry}
Semantically, we expect \code{notIf$\;$(False$\;$?$\;$True)}
to be equivalent to \code{True$\;$?$\;$False}.
However, with the memoization implementation given above,
we get the result \code{True$\;$?$\;$True}.
To see why this happens, let us look at the monadic transformation
of \code{notIf}.
\begin{haskell}
notIfC :: Curry (BoolC :-> BoolC)
notIfC = return (Func (\x ->
           let nx' = share x >>= \x' -> apply notC x'
           in share nx' >>= \nx -> x >>= \x' -> case x' of
                TrueC  -> nx
                FalseC -> nx ))
\end{haskell}
Remember that \code{x} has already been shared during the application
of \code{notIf} to \code{True ? False}.
Consequently, \code{x} is evaluated before \code{nx}
 since \code{x} is demanded by \code{notC}.
When the computation of \code{nx} is forced in either of the branches,
the value for \code{x} is memoized and can just be returned
for any subsequent computation without triggering any new non-determinism.
Thus, the current implementation of memoization
deduces that \code{nx} is deterministic and
the value of \code{nx} is memoized across branches as \code{TrueC}.
As a result, our implementation fails in situations where
a shared value (\code{nx}) depends on another shared value (\code{x})
that is forced earlier than the first one.

To fix this, we also memoize whether a value is deterministic
and inserted globally, or non-deterministic and inserted locally.
On reading a memoized value that was non-deterministic,
we simply advance the branch identifier.
Thus, a repeated sharing of the same non-deterministic value or a value
that directly depends on it will be assumed to be non-deterministic.
In other words, we now prevent global memoization of a computation
if it depends on a non-deterministic computation and
not just when the computation itself introduces non-determinism.
Therefore, we need to adapt the implementation.
The relevant changes are in both \code{case}-branches.
\begin{haskell}
case lookupTaskResult trRef b2 p2 of
  Nothing  -> do
    a <- ma
    MemoState b3 _ _ <- get
    let wasND = b2 /= b3
    let whereTo = if wasND then b3 else b1
    let addTR h = insertHeap whereTo (a, wasND) h
    unsafePerformIO (modifyIORef trRef addTR)
      `seq` return a
  Just (res, wasND) ->
    | wasND -> do
        i <- freshID
        let newParents = Set.insert b2 p2
        put (MemoState i newParents idSupply2)
        return res
    | otherwise ->
        return res
\end{haskell}

\subsection{Improving Performance of Sharing}
\label{sec:performance}

There is one major performance bottleneck with this implementation.
The problem is that \code{share} enters its argument into the memoization heap, even when a previous \code{share} has done so already.
Repeated sharing of a computation is unnecessary and expensive in both time and memory consumption, as each \code{share} introduces another indirection with an \code{IORef}.
Thus, an obvious performance optimization is to prevent repeated sharing of values as much as possible.
To achieve this, we use the following idea.

We prevent unnecessary insertions of \code{share} during
the monadic transformation whenever possible.
Any locally introduced variable (bound via let, case or lambda)
has already been shared by either the application or the let rule.
Thus, sharing these variables again when they are applied to a function
is unnecessary.
Consider the following example of a function that reverses
the order of elements in a list.
\begin{curry}
reverse :: [a] -> [a]
reverse []     = []
reverse (x:xs) = reverse xs ++ [x]
\end{curry}
Here, both \code{x} and \code{xs} have already been shared when they were originally applied to the list constructor \code{(:)}.
Thus, we can avoid sharing them explicitly.
To introduce this improvement, we replace
in the rules given in \cref{fig:transformation-expressions}
the definition of $alias$ by the new definition shown in
\cref{fig:improved-alias}.

\begin{figure*}
\begin{minipage}[c]{\textwidth}
  \begin{align*}
  \mcode{$alias(e_1, x, e_2)$}
    &\coloneqq
    \begin{cases}
      \mcode{let $x$ = $e_1$ in $e_2$} &
        \text{if}~e_1~\text{is a single, locally bound variable} \\
        \mcode{share $e_1$ $\bind$ $\lambda x$ $\rightarrow$ $e_2$} &
          \text{if}~e_1~\text{is a complex expression or a globally bound variable}  \\
      \end{cases}
      &\tag{Aliasing}
  \end{align*}
\end{minipage}
\caption{Improved $alias$ rule}
\label{fig:improved-alias}
\end{figure*}

\section{Extending the Monad}
\label{sec:extensions}

An important advantage of our monadic implementation is that
one can implement new features by modifying the monad without
changing the translation of source programs.
In this section, we will show extensions that are relevant
in contemporary functional logic languages.
The first extension concerns the addition of
free (logic) variables, which represent unknown values or possibly
infinite sets of values.
Free variables are required to support unification,
i.e., binding free variables to partially known values instead
of enumerating all their values.
Thus, we will show how unification can be implemented in our
monadic approach.
The second extension is encapsulated search,
i.e., a method to collect the results of non-deterministic
computations in some data structure.
Though this is known in logic programming for a long time (\code{findall}),
the interaction with demand-driven computations causes new challenges
which we will discuss.

\subsection{Free Variables}

Free variables are an important feature of logic programming.
They denote unknown values which are refined during a computation.
It is well known that
free variables can be replaced by value generators,
i.e., non-deterministic operations that yield the
(possibly infinite) set of values of a given type \citep{AntoyHanus06ICLP}.
For instance, a value generator \code{aBool} for Boolean values was defined
in \cref{ex:aBool}.
Similarly, a value generator for lists of Booleans
can be defined by
\begin{curry}
aBoolList ::  [Bool]
aBoolList = [] ? (aBool : aBoolList)
\end{curry}
Although the use of value generators for unknown values
is sufficient from a declarative point of view,
a crucial feature of logic programming is \emph{unification}
to compute with partial information, i.e., without completely
instantiating free variables.
For instance, if \code{x} and \code{y} are free variables,
the unification \code{x$\;$=:=$\;$y} is solved by binding
\code{x} to \code{y} (or vice versa) without instantiating
them to a value.
In contrast, if \code{x} and \code{y} are generators for an infinite
set of values, their evaluation might lead to an infinite search space.
In order to support an efficient implementation of unification,
we need to explicitly represent free variables during run time
so that we can implement bindings without generating values.

For this purpose, we use the approach taken by \citet{teegen21inversion}
and adapt it to our memoization monad.
Central to their approach is the idea that, if a computation demands
a free variable by using \code{(>>=)}, the variable gets narrowed,
i.e., it is partially instantiated.
This fits to our monadic compilation scheme:
whenever a value of some expression is demanded,
the bind operator \code{(>>=)} is used, i.e.,
to extract the scrutinee of a \code{case}-match
or to extract a function to be applied.\footnote{The operation \fcode{(>>=)}
is also used
on the result of a \fcode{share}, but this is irrelevant here as
\fcode{share} never yields a free variable.}
In both cases, we need an explicit value instead of a free variable
in order to continue the evaluation.

\citet{teegen21inversion} define a type class \code{Narrowable}, which
enumerates all constructors of the given type and fills their
arguments with new free variables.\footnote{Our implementation of
\fcode{Narrowable} is actually a bit simpler compared to
\citep{teegen21inversion}, but the reasons why that is possible
do not matter here.}

\begin{haskell}
class Narrowable a where
  narrow :: [Curry a]
\end{haskell}
Implementations for this class have to be generated for every transformed data type.
For example, the instance for the list type looks as follows.
\begin{haskell}
instance$\;\;$Narrowable a =>$~$Narrowable (ListC a)
 where
  narrow = [return NilC, ConsC <*> share free <*> share free]
\end{haskell}
Free variables are created using the function \haskellinline{free}
that will be introduced later.
We also need to extend our state with a heap to store the bindings
of free variables that have already been narrowed.
Since not all free variables have the same type,
our heap has to be untyped using existential data types
\citep{Perry2005Thesis} and \code{unsafeCoerce} as shown below.
Using unsafe features here will not cause issues at run time,
as in a type-correct program every logic variable has a unique identifier
and a single type.
\begin{haskell}
data Untyped = forall a. Untyped a$\listline$
typed :: Untyped -> a
typed (Untyped x) = unsafeCoerce x$\listline$
data FreeState = FreeState {
    branchID  :: ID,
    parentIDs :: Set ID,
    varHeap   :: Heap Untyped,
    idSupply  :: IORef ID
  }
\end{haskell}
Now we can define the new type for our memoization monad
extended with free variables.
Apart from the state, it differs from our previous monad by having
the type \code{(FLVal$\;$a)} instead of a plain \code{a} in its result.
This new type differentiates between values and variables,
with the latter one containing a \code{Narrowable} constraint
so that we are guaranteed to be able to narrow a variable
that we encounter during evaluation:
\begin{haskell}
data FLVal a = Val a
             | Narrowable a => Var ID$\listline$
\end{haskell}
Thus, the state of computations with free variables is defined
as\footnote{Note that we cannot derive required instances
(e.g., \code{Monad}) automatically anymore.}
\begin{haskell}
newtype CurryFree a = CF {
    unCF :: StateT FreeState Tree (FLVal a)
  }
\end{haskell}
Now we can show the implementation of
\haskellinline{free} that we mentioned earlier.
It uses the new type \code{FLVal} and provides a fresh identifier
of some type constraint to \code{Narrowable}.
Here we use the \code{ID} type to uniquely identify free variables
in addition to their use as branch identifiers.
\begin{haskell}
free :: Narrowable a => CurryFree a
free = CF <> do
  i <- freshID
  return (Var i)
\end{haskell}
Before we define the \code{Monad} instance for the \code{CurryFree} type,
we introduce another helper function for the instantiation of free variables.
This instantiation starts by generating a transformed value for each constructor
of the corresponding type using \code{Narrowable}.
For each of those values, we spawn a new computation that inserts a binding for the instantiated variable into the heap and
updates the branch identifier accordingly.
Thus, the \code{varHeap} in every branch contains a mapping from
variable identifiers to monadic computations that have already been shared.
We then combine the computations for each constructor
by non-deterministic choices using \code{msum}.
It is imperative that we first bind the corresponding variable and
then write it on the heap so that repeated lookups of the same variable
also share the same identifiers generated by calls to \code{free}
in the \code{Narrowable} instance.
Otherwise, two computations in the same branch could compute
different identifiers for the free variables in the arguments
of a narrowed constructor, leading to unsound results.
We do, however, write only monadic values on the heap so that we can also put lazy computations on the heap. Thus, we wrap our instantiated value in a \code{CF (return ..)}.
\begin{haskell}
instantiate :: Narrowable a => ID -> CurryFree a
instantiate vid = CF <> do
  st <- get
  msum (map (update st) narrow)
 where
  update (FreeState i ps h suppl) x = unC <> do
    i' <- freshID
    sharedX <- x
    let h' = insertHeap vid (Untyped (CF (return sharedX))) h
    put (FreeState i' (Set.insert i ps) h' suppl)
    return sharedX
\end{haskell}
Now we can implement the \code{Monad} instance of \code{CurryFree}.
The only interesting bit is that we make a case distinction
on the result of the first argument so that we can check
whether it is a variable or a value.
Recall that we have to instantiate free variables when
they are pattern matched in the source program,
which corresponds to a \code{(>>=)} in the monadic translation.
In case the result is a variable that is yet unbound,
we instantiate that variable before we continue with the computation.
Otherwise, we can apply the continuation from the second argument.
\begin{haskell}
instance Monad CurryFree where
  CF ma >>= f = CF <> do
    fla <- ma
    FreeState _ _ h _ <- get
    unCF <> case fla of
      Var i -> case lookupHeap i h of
                 Nothing -> instantiate i >>= f
                 Just x  -> typed x >>= f
      Val x -> f x
\end{haskell}

\subsection{Monadic Unification}
\label{sec:unification}

With the extension of our monad to represent free variables,
we are able to implement a standard unification procedure
by inserting variable bindings directly onto the heap
rather than instantiating these variables.

In the following, we show an implementation of a Prolog-like unification
algorithm based on the monad described in the previous section.
The important issue is the distinction between free variables
and values (see type \code{FLVal} defined above).
If free variables are unified, a binding between these variables
is stored onto the heap instead of instantiating them.

We start by defining a type class \code{Unifiable}
which contains the operation \code{unify}.
\begin{haskell}
class Unifiable a where
  unify :: a -> a -> CurryFree BoolC
\end{haskell}
Based on \code{unify}, we define a generic operation \code{unifyC}
which checks whether the arguments are free variables or values.
Before the check, we have to follow any chain of variables with \code{deref}
since it can happen that a variable is bound to another one.
If both arguments are unbound free variables,
we bind one variable to the other without instantiating any of them.
In the case that only one of the arguments is a variable, we instantiate it and unify recursively.
Whenever both arguments are already instantiated, we simply unify their values.
\begin{haskell}
unifyC :: Unifiable a
       => CurryFree a -> CurryFree a -> CurryFree BoolC
unifyC ma mb = CF <> do
  fla <- deref ma
  flb <- deref mb
  unCF <> case (fla, flb) of
    (Var i, Var j) -> do
      FreeState b ps h suppl <- get
      let h' = insertHeap i (Untyped (CF (return (Var j)))) h
      put (FreeState b ps h' suppl)
      return TrueC
    (Var i, Val y) -> instantiate i >>= unify y
    (Val x, Var j) -> instantiate j >>= unify x
    (Val x, Val y) -> unify x y
  where
    deref (CF m) = do
      fl <- m
      case fl of
        Var i -> get >>= \ (FreeState _ _ h _) ->
                 case lookupHeap i h of
                   Just x  -> deref (typed x)
                   Nothing -> return (Var i)
        Val x -> return (Val x)
\end{haskell}
The instances of \code{Unifiable} can be schematically generated
together with the transformation of data types.
Note that \code{unify} either returns \code{TrueC},
if both arguments are unifiable, or fails.
The instances for base types are easy, as shown for the type of Booleans.
\begin{haskell}
instance Unifiable BoolC where
  unify FalseC FalseC = return TrueC
  unify TrueC  TrueC  = return TrueC
  unify _      _      = mzero
\end{haskell}
Structured types are unified by pairwise unifying the arguments
of identical data constructors, as shown for the list data type.
\begin{haskell}
instance Unifiable a => Unifiable (ListC a)
 where
  unify NilC         NilC         = return TrueC
  unify (ConsC x xs) (ConsC y ys) = (&&) <§>
                                    unifyC x y <*> unifyC xs ys
  unify _            _            = mzero
\end{haskell}
Finally, Curry's unification operator, as introduced in \cref{sec:flp},
can be implemented as follows.
\begin{haskell}
(=:=) :: Unifiable a => CurryFree (a :-> a :-> BoolC)
(=:=) = return <> Func <> \x ->
          return <> Func <> \y -> unifyC x y
\end{haskell}
The explicit instances of \code{Unifiable} for each data type
might look cumbersome but, actually, it is a feature of our approach.
All these instances can be schematically generated
by a compiler.
More important, it is not reasonable to have instances for all types.
\citet{HanusTeegen20} proposed the introduction of a type class \code{Data}
which supports value generators, strict equality, and unification
for data types.
Since functional types have no \code{Data} instances,
the operators supported by the \code{Data} class are overloaded
so that a different implementation is required for each data type,
as in our approach.

\subsection{Encapsulated Search}
\label{sec:encapsulated-search}

Non-deterministic programming with built-in search
supports a compact and elegant programming style,
as known from logic programming and also used
in functional logic programming \citep{AntoyHanus02FLOPS}.
In application programs, it is necessary to encapsulate
non-deterministic computations so that the outcomes
are collected in some data structure, e.g., to print the
results in a deterministic manner or to select a result
according to some criterion.
Prolog offers various operations for this purpose,
like \code{findall} and similar predicates.
However, a direct transfer of such predicates to
a language with a demand-driven evaluation strategy is problematic.
For instance, \code{findall} always computes all solutions
so that it cannot be applied to infinite search spaces.
If an encapsulation operator would return a possibly
infinite list or tree, lazy evaluation allows to process
part of this structure in a finite computation.
Therefore, proposals for encapsulated search in functional logic languages
try to support demand-driven evaluation of non-deterministic computations,
e.g., \citep{AntoyBrassel07,AntoyHanus09,ChristiansenHanusReckSeidel13PPDP}.
For instance, if \code{allValues$\;e$} denotes the list of all
values of the expression $e$, one can test whether $e$ has no value
by \code{null$\;$(allValues$\;e$)}.
Thanks to lazy evaluation, one gets a Boolean result
even if $e$ has infinitely many values.

Unfortunately, the combination of encapsulated search
and demand-driven evaluation causes new challenges.
A potential problem occurs if the encapsulated expression $e$
shares subexpressions introduced outside the
encapsulation operator.
For instance, consider the expression
\begin{curry}
let b = False ? True in allValues (not b)
\end{curry}
The obvious question is whether the non-determinism of \code{b}
should be encapsulated by \code{allValues} or not.
There is no clear answer but there are different views.
\emph{Strong encapsulation}, inspired by Prolog's \code{findall},
requires encapsulating all non-determinism accessible from the
encapsulated expression so that this expression yields
the single list \code{[True,False]}.\footnote{One could also return
a multi-set instead of a list in order to avoid enforcing
an ordering on the solutions.}
Unfortunately, the results of strongly encapsulating search operators
depend on the evaluation strategy. If we consider the expression
\begin{curry}
let b = False ? True in (allValues (not b), b)
\end{curry}
the computed pairs of values depend on the order of evaluating
the components of the pair. If they are evaluated from left to right,
we obtain the list of values
\begin{curry}
[([True,False], False), ([False,True], True)]
\end{curry}
With a right-to-left evaluation, we obtain
\code{[([True], False), ([False], True)]} since, due to sharing,
the already evaluated values of \code{b} are encapsulated.
These and more pitfalls of strong encapsulation are discussed
\citep{BrasselHanusHuch04JFLP}.

One can avoid these pitfalls by a clear distinction between
expressions evaluated \emph{inside} an encapsulation operator,
i.e., where all values of these expressions are collected,
and \emph{outside}, where different values lead to different results.
For this purpose, \citet{AntoyHanus09} proposed \emph{set functions}
which encapsulate the non-determinism caused by a function
but do not encapsulate the non-determinism of arguments.
It has been shown that set functions are a strategy-independent
notion of encapsulating search in demand-driven non-deterministic languages.

Instead of discussing more details about various search operators,
we want to show how we can implement strong encapsulation
(other search operators can be based on this
by distinguishing the subexpressions to be encapsulated).
In the following, we will see a further advantage of our monadic approach.
Instead of adapting the compilation scheme to support
encapsulated search, as done in other Curry compilers
(see \cref{sec:related-work}), we use the same monadic target programs
but put all the extensions inside the implementation of the monad.

A crucial step for strong encapsulation in a lazy language
is reducing an arbitrary expression to a fully-evaluated,
constructor-based value.
To do this for arbitrary data types,
we define a type class \haskellinline{NormalForm} and an additional operation \haskellinline{normalFormCurry}.\footnote{Our
actual compiler uses a version of the given class
that is overloaded in the recursive call.This enables the definition
of a normal form computation that instantiates free variables and
one that does not}.
\begin{haskell}
class NormalForm a where
  normalForm :: a -> Curry a

normalFormCurry :: NormalForm a => Curry a -> Curry a
normalFormCurry ma = ma >>= normalForm
\end{haskell}
Since the arguments of an arbitrary data type might contain non-deterministic computations, the result of \haskellinline{normalForm} can be non-deterministic.
Conceptually, the operation \enquote{pulls} the non-determinism to the outside of a data structure, which is why this is also known as pull-tabbing.

For our list data type, an instance looks as follows.
\begin{haskell}
instance NormalForm a => NormalForm (ListC a) where
  normalForm NilC         = return NilC
  normalForm (ConsC x xs) = do
    x' <- normalForm x
    xs' <- normalForm xs
    return (ConsC (return x') (return xs'))
\end{haskell}
Using this type class, \curryinline{allValues} can be defined by
evaluating the normal form of the encapsulated expression with the
current state, collecting all results using some tree traversal (here:
depth-first), and converting the Haskell-list into the internal
Curry-list representation.

\begin{haskell}
allValues :: NormalForm a => Curry a -> Curry (ListC a)
allValues ma = Curry <> get >>= \state ->
  let tree = evalStateT (unCurry (normalFormCurry ma)) state
  in return (toCurryList (dfs tree))
\end{haskell}
The depth-first tree traversal and list conversion are simply defined as:
\begin{haskell}
dfs :: Tree a -> [a]
dfs Empty      = []
dfs (Leaf x)   = [x]
dfs (Node a b) = dfs a ++ dfs b$\listline$
toCurryList :: [a] -> ListC a
toCurryList []     = NilC
toCurryList (x:xs) = ConsC (return x) (return (toCurryList xs))
\end{haskell}
An advantage of the tree representation for non-determinism is that
we can replace the depth-first tree traversal by other ones,
e.g., breadth-first or iterative deepening.
An interesting alternative is an implementation
of a \textit{fair search} \citep{BoehmHanusTeegen21}
as described in the following.

\subsection{Fair Search}
Consider the following non-deterministic definition.
\begin{curry}
sometimesLoops :: Bool
sometimesLoops = loop ? True ? loop
  where loop = loop
\end{curry}
Conceptually, \curryinline{sometimesLoops} has exactly one result.
However, both a depth-first search and a breadth-first search will
loop forever without ever finding the value \curryinline{True} due to
the non-terminating computation in the left and/or right branch.
As an alternative, a \textit{fair search} strategy should ensure that
all results are computed in a finite amount of time, even in the presence
of non-terminating computations in non-deterministic branches.

One can implement a fair search strategy in Haskell by traversing
non-deterministic branches of the search tree in concurrent threads
where leaf values are written into an answer channel.
A basic version of this idea can be implemented with the
concurrency features of Haskell \citep{PeytonJonesGordonFinne96POPL}.
\begin{haskell}
fsBasic :: Tree a -> [a]
fsBasic t = unsafePerformIO <> do
  ch <- newChan
  let go t' = case t' of
                Empty    -> return ()
                Leaf x   -> writeChan ch x
                Node l r -> do
                  _ <- forkFinally (go l) (\ _ -> return ())
                  _ <- forkFinally (go r) (\ _ -> return ())
                  return ()
  _ <- fork(go t)
  getChanContents ch
\end{haskell}
This implementation performs the following steps:
\begin{enumerate}
  \item Create a new channel \code{ch} for communicating result values.
  \item Define a recursive function \code{go} to traverse the tree where
    \begin{itemize}
    \item leaf values are directly written in the channel \code{ch}, and
    \item concurrent threads are spawned for left and right subtrees.
    \end{itemize}
  \item Run \code{go} concurrently and collect results by using
        \code{getChanContents}.
\end{enumerate}
While this fair search will spawn a huge amount of threads, these are
all relatively lightweight and directly managed by GHC's run-time system.

A disadvantage of this simple approach is that the channel stays
open, even when all threads have already completed their work.
To determine when all threads have finished their work, we use
Haskell's mutable variables \haskellinline{MVar}
\citep{PeytonJonesGordonFinne96POPL} as a semaphore to
compensate the lack of a \textit{join} operation on threads.
If a thread tries to read an empty \haskellinline{MVar},
it suspends until it is eventually filled by other threads.
When all threads have completed their work,
the outermost thread will write a \haskellinline{Nothing} value
into the channel to signal completion.
Since \haskellinline{getChanContents} returns the results lazily, taking
values until we find \haskellinline{Nothing} is also lazy enough.

\begin{haskell}
fs :: Tree a -> [a]
fs t = unsafePerformIO <> do
  ch <- newChan
  let go t' = case t' of
                Empty    -> return ()
                Leaf x   -> writeChan ch (Just x)
                Node l r -> do
                  (mR, mL) <- (,) <§> newEmptyMVar <*> newEmptyMVar
                  _ <- forkFinally (go l) (\_ -> putMVar mL ())
                  _ <- forkFinally (go r) (\_ -> putMVar mR ())
                  takeMVar mL >> takeMVar mR
  _ <- forkFinally (go t) (\_ -> writeChan ch Nothing)
  catMaybes . takeWhile isJust <§> getChanContents ch
\end{haskell}
One further optimization implemented in our compiler is the use
of \textit{finalizers} to \enquote{hook} onto the garbage collector
based on the concepts described by \citet{PeytonJones2000Finalize}.
When the garbage collector cleans up the result list of our fair search,
we know that no more elements will be taken out of the channel.
Thus, we can stop all threads that are still active.
For instance, if we demand only a single value from a search tree
with more non-deterministic branches, the remaining threads
are automatically terminated after getting a value and garbage collecting
the list of possible values.

\subsection{Further Extensions}

As mentioned in \cref{sec:flp}, there are still some other extensions
that have been proposed for high-level declarative programming.
The most relevant are \emph{functional patterns} \citep{AntoyHanus05LOPSTR}
where one can use defined functions in addition to data constructors
in patterns.
Functional patterns support a compact programming style
since a functional pattern abbreviates a (possibly infinite) set
of standard patterns.
This allows for deep pattern matching which is exploited
in \citep{Hanus11ICLP} to process XML structures.

For instance, the operation \code{last}, shown in \cref{example:last} in a classical functional logic programming style, can be defined with a functional pattern by
\begin{curry}
last (_ ++ [x]) = x
\end{curry}
Although this is basically the specification of the property
to be satisfied by \code{last}, it is also executable
if functional patterns are supported and even yield
results if the list contains failing computations before the last element.
Conceptually, a functional pattern denotes all regular patterns
to which it can be evaluated.
In our example, these are all lists with \code{x} as a final element.
\citet{AntoyHanus05LOPSTR} showed that functional patterns can be
implemented by a non-strict unification procedure where
pattern variables are not fully evaluated, in contrast to the
operator \code{unify} described in \cref{sec:unification}.
Since the necessary changes are limited, we skip a detailed
description of them.

The full implementation of our monad and compiler with all extensions
described above is available in our
GitHub repository {\small \url{https://github.com/Ziharrk/kmcc}}.

\section{Optimizations for Deterministic Expressions}
\label{sec:optimize}

Not all operations defined in a Curry program make use of non-determinism
and other logic features.
In fact, only certain parts of a program are written in a
non-deterministic style.
These parts are usually encapsulated (see \cref{sec:encapsulated-search})
and their results are then further processed by purely functional
or IO-based code.
To optimize the performance of our compilation model,
we can identify these deterministic parts and essentially keep them as-is in the target Haskell code.
As an example, consider the following definitions.
\begin{curry}
permutations123 :: [Int]
permutations123 = perm [1,2,3]$\listline$
permutationLengths :: [Int]
permutationLengths = map length (allValues permutations123)
\end{curry}
The functions \curryinline{map} and \curryinline{length} are both
deterministic.
Thus, computing the size of a list of permuted values should only require
a non-deterministic computation for the permutation itself,
whereas the remaining part of the computation is purely functional.

Deterministic parts of a program can be identified
with a simple static analysis using a fixed-point computation.
In the following, we will focus on using this information
to optimize the compilation of deterministic parts of a program.

\subsection{Deterministic Sub-Computations}

In \cref{sec:lifting} we have shown that data types are transformed into a variant where each constructor argument is a monadic computation.
However, in order to keep deterministic parts of a program as-is,
we need to keep a deterministic copy of each function
together with the original data types.
When the result of a deterministic function is passed
into a non-deterministic operation,
we have to convert the plain representation into the monadic representation
of the given data type.
These conversions can be easily generated for each data type.
In our implementation, they are part of a type class \haskellinline{FromHs}.
Since our target language Haskell is typed, we also need a connection between these data types on the type level.
For this purpose, we use an injective type family
\citep{Eisenberg14Closed}
which maps each Curry type to its Haskell representation.
The corresponding class and type family definition as well as an example instance for lists are shown below.
\begin{haskell}
type family HsEquivalent (a :: k) = (b :: k) | b -> a$\listline$
class FromHs a where
  from :: HsEquivalent a -> a$\listline$
fromHs :: HsEquivalent a -> Curry a
fromHs = return . from$\listline$
type instance HsEquivalent ListC = List
instance FromHs a => FromHs (ListC a) where
  from Nil         = NilC
  from (Cons x xs) = ConsC (fromHs x) (fromHs xs)
\end{haskell}
With this conversion, the optimized transformation of the function
\haskellinline{permutation123} from the example above can be implemented
as follows.
\begin{haskell}
permutations123 :: Curry (ListC Int)
permutations123 = permC (fromHs [1,2,3])
\end{haskell}
There is just one catch with this optimization in the general setting:
functions as arguments.
If a deterministic function is passed into a non-deterministic one,
we need to convert the function into its monadic representation.
However, this is not possible in general:
the converted function has to operate on monadic values
but the function we are trying to convert operates on pure values.
One could encapsulate the non-deterministic argument values and map
the original function over all these values,
but this approach is too strict in certain cases.
Thus, our optimization needs to ensure that no deterministic function
is passed as an argument to a non-deterministic one.
This can be a bit challenging to implement
since a function could be \enquote{hidden} inside a data type.
In any of these cases, we simply fall back to the non-deterministic
variants of the involved functions and data types.

A further complication comes from Curry's demand-driven evaluation strategy.
A deterministic function might trigger non-determinism when
forcing the evaluation of one of its arguments.
Thus, each argument of a deterministic function
needs to be deterministic so that the optimized version is applicable.

Due to these complications, a purely static approach
would often use non-deterministic variants of functions
where it is not necessary at run time.
Without the ability to detect deterministic values at run time,
once we enter the realm of non-determinism, there is no turning back.

\subsection{Marking Deterministic Values}

In order to avoid the problems mentioned above,
we can mark a value as deterministic at run time.
To do this, we extend each data type by a new constructor
with a single argument that contains the deterministic variant of the given type.
The updated transformation rule for data types is shown
in \cref{fig:transformation-data-opt}.
For the \haskellinline{List} type, the final definition of
its non-deterministic variant is as follows.
\begin{haskell}
data ListC a = NilC
             | ConsC (Curry a) (Curry (ListC a))
             | ListC# (HsEquivalent (List a))
\end{haskell}
\begin{figure*}
  \begin{minipage}[c]{\textwidth}
    \begin{align*}
    \mcode{$\llbracket$\textbf{data} $D$ $\alpha_1 \ldots \alpha_n$ = $C_1$ | $\ldots$ | $C_n$ $\rrbracket^d$}
      &\coloneqq \mcode{\textbf{data} \rename($D$) $\alpha_1 \ldots \alpha_n$ = $\llbracket C_1 \rrbracket^c$ | $\ldots$ | $\llbracket C_n \rrbracket^c$} \\
      & \mcode{\ \ \ \ | $D\#$ (HsEquivalent ($D$ $\alpha_1 \ldots \alpha_n$))}\tag{Data type}
    \end{align*}
  \end{minipage}
  \caption{Revised data type transformation $\llbracket \circ \rrbracket^d$}
  \label{fig:transformation-data-opt}
\end{figure*}
In some cases it can be convenient to have a dual function to
\haskellinline{fromHs} which receives a value in its monadic representation
and returns a computation that yields the deterministic representation
of the given value.
Our compiler uses this function for some foreign function implementations
but it is not further relevant to this paper.

Since we added another constructor to each data type,
our previous transformation for case expressions becomes incomplete.
We need to add a case for the new constructor where we exploit the fact
that the argument is known to be deterministic.
Since it is easier to define, we duplicate each branch of the case expression,
where the new branch pattern includes the new constructor.
The revised transformation rule for branches of case expressions
is given in \cref{fig:transformation-case-opt}.
It omits the right-hand side of the new branch
since it is hard to capture the decision of whether and how to use
the deterministic variant of a function in a compact way.
The decision is based on the used in-scope variables and
their determinism status.
An example of the new transformation is shown below for the
\haskellinline{length} function, where \haskellinline{succ} is
the increment function.
\begin{haskell}
lengthC :: Curry (ListC a :-> IntC)
lengthC = return (Func (\xs ->
  case xs of
    NilC               -> return 0
    ListC# Nil         -> return 0
    ConsC _ xs         -> share (apply lengthC xs) >>= apply succC
    ListC# (Cons _ xs) -> return (succ (length xs)) ))
\end{haskell}
Note that one of the branches exclusively uses the deterministic function
for the recursive call.
If we were to assume that \haskellinline{length} is non-deterministic, the last branch would look as follows.
\begin{haskell}
    ListC# (Cons _ xs') -> let xs = fromHs xs' in
                           succC >>= \f -> f (lengthC xs)
\end{haskell}

\begin{figure*}
  \begin{minipage}[c]{\textwidth}
    \begin{align*}
    \mcode{$\llbracket C~x_1 \ldots x_n \rightarrow e \rrbracket^b$}
      &\coloneqq \mcode{\rename($C$) $x_1 \ldots x_n \rightarrow$}\ \mcode{$\llbracket e \rrbracket^e$;} & \\
      &\ \ \ \ \ \; \mcode{$D\#$ ($C$ $x_1 \ldots x_n) \rightarrow$}\ \mcode{(..)} \tag{Case Branch}
    \end{align*}
  \end{minipage}
  \caption{Revised transformation rule for branches of case expressions $\llbracket \circ \rrbracket^b$}
  \label{fig:transformation-case-opt}
\end{figure*}

One might notice that doubling the number of branches can lead
to an exponential increase in the size of the generated code
in case of deeply nested case expressions.
Actually, this turned out to be practically relevant
since there are programs where the flat form yields
large case expressions (e.g., pattern matching on strings).
This can be avoided by transforming programs so that there is
at most one case expression directly at the root of the function.
This form of programs, which are also called \emph{uniform},
can be obtained by moving nested case expressions
into new functions \citep{MorenoEtAl90ALP}.
The transformation into uniform programs is also used
in other Curry compilers \citep{BrasselHanusPeemoellerReck11,Hanus25PAKCS}.
With uniform programs,
we do not need to double the number of branches in case expressions
but only add a single one for the new constructor where
we call the appropriate version of the same function we are already in.
This allows us to still \enquote{switch} to the deterministic world
without an exponential code increase.

One nice additional benefit of the new constructor is that we can use it to defer the conversion of a value in the \haskellinline{FromHs} class.
Instead of converting a value to its monadic representation immediately, we can just wrap it in the new constructor.
This way, we can still pass the value or its sub-components to the deterministic variant of a function.

\section{Evaluation}
\label{sec:evaluation}

The objective of this work is to create a high-level
memoization implementation for Curry based on a purely functional
interface.
This implementation is intended to support sharing of values
to avoid repeated pull-tab steps and to allow for sharing
of deterministic computations even across non-determinism.

Due to the monadic abstraction, the implementation of our compiler
allowed for a clean separation between the run-time system
(i.e., the monad and \code{share} implementation)
and the monadic transformation.
The full compiler implementation that we have now shows that
the code size is smaller
(thus, better maintainable) and more modular
than KiCS2 \citep{BrasselHanusPeemoellerReck11}
which also compiles into Haskell.

In order to evaluate the overall efficiency of our high-level
implementation, we compare it on various benchmark programs\footnote{The
benchmarks are available in our GitHub repository.}
with three other major Curry implementations.
PAKCS \citep{Hanus25PAKCS} compiles to Prolog (SWI-Prolog 9.0.4)
so that its search strategy is based on backtracking.
KiCS2 \citep{BrasselHanusPeemoellerReck11} compiles to Haskell (GHC 9.4.5)
and implements non-determinism by pure pull-tabbing without memoization.
Curry2Go \citep{BoehmHanusTeegen21} compiles to Go (1.19.3)
and uses an imperative implementation of memoization.
All benchmarks were executed on a Linux machine
running Debian 12 with an Intel Core i7-7700K (4.2GHz) processor
with eight cores.
We measured the average elapsed time (in seconds) of three runs using
the \code{time} command and the executables generated
by the respective compilers.

\begin{figure*}
\begin{minipage}[c]{0.9\textwidth}
\begin{tabular}[c] {
  | l |
  | r
  | r
  | r
  | r
  | r | }
\hline
Program & \multicolumn{1}{c|}{Monadic MPT} & \multicolumn{1}{c|}{Monadic MPT + det. optim.} & \multicolumn{1}{c|}{KiCS2} & \multicolumn{1}{c|}{PAKCS} & \multicolumn{1}{c|}{Curry2Go} \\
\hline
\hline
\texttt{addNum5}      &   2.86 &                      2.82 &  4.36 & \cellcolor{green!30}   0.33 &  0.43 \\
\texttt{addNum10}     &   4.29 &                      4.25 & 13.70 & \cellcolor{green!30}   0.41 &  0.68 \\
\texttt{select50}     &   0.08 & \cellcolor{green!30} 0.02 &  0.34 &                        0.26 &  0.05 \\
\texttt{select100}    &   0.57 & \cellcolor{green!30} 0.04 &  5.27 &                        0.27 &  0.07 \\
\texttt{select150}    &   1.90 & \cellcolor{green!30} 0.09 & 28.43 &                        0.30 &  0.18 \\
\texttt{yesSharingND} &   1.33 & \cellcolor{green!30} 0.02 &  0.42 &                       33.65 &  4.62 \\
\texttt{noSharingND}  &   2.58 & \cellcolor{green!30} 0.02 &  0.78 &                       34.07 &  2.97 \\
\texttt{permSort}     &   2.55 & \cellcolor{green!30} 2.46 &  2.80 &                       10.30 &  5.96 \\
\texttt{sortPrimes}   &   7.67 & \cellcolor{green!30} 0.01 &  0.03 &                       98.56 &  1.02 \\
\texttt{naiveReverse} &   3.52 & \cellcolor{green!30} 0.14 &  0.20 &                        6.46 &  1.15 \\
\texttt{queens10}     & 221.26 & \cellcolor{green!30} 0.09 &  0.45 &                      185.31 & 27.41 \\
\hline
\end{tabular}
\end{minipage}
\caption{Timings (in seconds) of various programs evaluated with different compilers, green marks best time}
\label{fig:benchmarks}
\end{figure*}

\cref{fig:benchmarks} shows the timings for various programs.
The first seven benchmarks test the properties
that our implementation was designed for,
while the last two benchmarks are purely deterministic programs.
The remaining two benchmarks test expensive non-deterministic computations.
Times for our approach are both \enquote{Monadic MPT} columns,
where the latter one integrates the optimization
for fully deterministic sub-computations from \cref{sec:optimize}.
Both are performed with a depth-first search strategy,
GHC optimizations (\texttt{-O1}) and its single-threaded RTS.
We will focus on the non-optimized version of \enquote{Monadic MPT} first
and discuss the optimized implementation at the end of this section.
The benchmarks are based on the ones used by \citet{BoehmHanusTeegen21}
to evaluate memoized pull-tabbing.

Compared to our initial prototype which was based on a direct integration
into GHC using a compiler plugin \citep{HanusProttTeegen22},
the unoptimized implementation of our compiler lost a lot of performance.
This seems to be the case for benchmarks with functions
that are \enquote{more strict} in their arguments, suggesting that
GHC's demand analysis was more effective for the plugin implementation.
However, our optimized implementation reaches
the same performance as the prototype for all but one benchmark.

The first five benchmarks check memoization, i.e., the avoidance
of repeated pull-tabbing.
The contrived programs \code{addNum5} and \code{addNum10}
test this property by non-deterministically generating
a number \code{x} between \code{0} and \code{2000}
(inclusive, see \code{someNum} below)
and adding \code{x} five and ten times to itself, respectively.
\begin{curry}
someNum :: Int -> Int
someNum n | n <= 0    = 0
          | otherwise = n ? someNum (n-1)$\listline$
addNum5 :: Int -> Int
addNum5 n = let x = someNum n in x+x+x+x+x
\end{curry}
With memoized pull-tabbing,
the choice for \code{x} should be made only once and
not on each addition of \code{x}.
Looking at the times in \cref{fig:benchmarks},
our implementation requires for \code{addNum10}
less than double the time of \code{addNum5}.
This is in contrast to KiCS2 which does not use
memoization to implement pull-tabbing.
Compared to our initial prototype,
this is the only benchmark where our optimized implementation
does not perform with the best timings.
We attribute this to the fact that the integration of the prototype
into GHC allowed for more aggressive optimizations,
especially for numeric-heavy computations.
It should be noted that PAKCS and Curry2Go, which are faster on this
benchmark, use a direct representation of integers
which do not support searching on integer variables (similarly to Prolog),
whereas our implementation allows for searching and
constraint solving on integers via Z3 \citep{deMouraBjorner08}.

The next three benchmarks \code{select$n$} non-deterministically
select an element in a list of length $n$ and sums up the element
and the list without the selected element.
They also demonstrate the advantage of memoized pull-tabbing
compared to the \enquote{raw} implementation of pull-tabbing in KiCS2.

The benchmarks \code{yesSharingND} and \code{noSharingND}
use the \code{primes} example from \cref{ex:primes}
to create an expensive deterministic computation
that yields the 800th prime number.
\begin{curry}
prime800 :: Int
prime800 = primes !! 799
\end{curry}
This computation is used in the following tests,
where the first one shares the value of \code{prime800}
through a \code{let}-binding across the non-deterministic choice
and the second one does not.\footnote{Note that top-level declarations
are always operations in Curry. Their results are never shared.}
\begin{curry}
yesSharingND :: Int
yesSharingND = let p = prime800 in p ? p$\listline$
noSharingND :: Int
noSharingND = prime800 ? prime800
\end{curry}
Since \code{noSharingND} takes approximately twice as long
as \code{yesSharingND},
we can conclude that our implementation shares deterministic values
across non-determinism.

The final benchmarks are taken from \citep{HanusTeegen21}.
\code{permSort} sorts a list of 13 elements by non-deterministically
generating all permutations and keeping the sorted one.
\code{sortPrimes} generates prime numbers as shown above and
sorts a list of four of them using permutation sort.
This benchmark mixes determinism and non-determinism.
The large execution time of PAKCS is due to the fact
that it does not implement sharing across non-determinism
so that the prime numbers are recomputed in each permutation.
\code{naiveReverse} is a simple deterministic example
of the quadratic algorithm to reverse a list of 4096 elements, and
\code{queens10} computes the number of safe positions to put 10 queens
on a 10 $\times$ 10 chessboard using Peano numbers.

These benchmarks indicate that, for purely functional programs,
our non-optimized implementation is faster than PAKCS and sometimes Curry2Go
but slower than KiCS2.
For non-deterministic programs, the results depend on the exact benchmark.
Since our implementation optimizes for programs such as \code{addNum10}
or \code{select150} where avoiding repeated pull-tabbing is relevant,
we are predictably faster than KiCS2.
PAKCS is sometimes faster since it uses backtracking in Prolog
instead of pull-tabbing.

Unfortunately, the non-optimized version is slower than Curry2Go
and much slower than KiCS2 in the mixed benchmark \code{sortPrimes}.
While this may seem surprising at first, the speed of KiCS2 stems
from the fact that it also optimizes deterministic operations,
even when deterministic functions are applied to potentially
non-deter\-ministic arguments.
This is actually very beneficial in the implementation of \code{sortPrimes}
and for the two deterministic benchmarks as well.
Without this optimization,
KiCS2 is about as slow as our unoptimized approach.

Using the optimizations for deterministic computations presented in
\cref{sec:optimize}, we can achieve a similar performance to KiCS2 and
sometimes even outperform it in the purely functional benchmarks.
This is due to the fact that for those functions our compiler
(together with GHC) essentially generates the same code as GHC would
have generated if we had compiled the benchmark as a Haskell program.
However, since our implementation fundamentally differs from the one
in KiCS2, it is still possible to construct programs where KiCS2
outperforms our compiler.

\section{Related Work}
\label{sec:related-work}

There are two different groups of related works which we discuss
in the following subsections.

\subsection{Other Compilers for Curry}
While we have already mentioned other Curry compilers throughout the paper,
here we summarize similarities and differences between them and our approach.

The compiler most similar to our work is KiCS2
\citep{BrasselHanusPeemoellerReck11}.
It compiles to Haskell and the evaluation strategy is based on pull-tabbing
as well.
However, KiCS2 does not support memoization.
While its model of Curry programs is not directly monadic,
it bears some resemblance.
Instead of having an explicit effect data type like \code{Tree}
in our approach, the structure is basically inlined by augmenting every data type with a constructor for non-deterministic choices and for failure.
Instead of using a state monad to pass around information,
KiCS2 augments every function with corresponding parameters.
This allows better optimization but
results in more complex and less maintainable code.
For instance, the implementation of set functions in KiCS2
requires a modification in the compilation scheme by adding
an additional argument (the encapsulation level)
to each function \citep{ChristiansenHanusReckSeidel13PPDP}.

\citet{Prott23Embedding} uses the same basic idea of a monadic
transformation to augment the GHC with different semantics,
including one with Curry-style non-determinism.
However, that implementation lacks the performance improvements and
concentrates on the integration into the Haskell compiler itself.

Other related compilers are Curry2Go \citep{BoehmHanusTeegen21}
and its Julia-based predecessor \citep{HanusTeegen21}.
These compilers introduced and refined the memoization approach
to pull-tabbing but both used an imperative language as their back end.
Thus, the compilation scheme and their modeling of Curry
are substantially different.

Sprite \citep{AntoyJost16} compiles Curry to LLVM
in order to generate efficient target code.
It is also based on pull-tabbing but does not implement memoization.
We could not include a comparison in our benchmarks since
a working implementation is not available.
The results published in \citep{AntoyJost16}
indicate that the performance of Sprite behaves similarly to that of KiCS2.

The PAKCS compiler from Curry to Prolog \citep{Hanus25PAKCS}
is very different from our approach.
Due to the use of Prolog as a target language for compilation,
PAKCS inherits the incompleteness of backtracking as a search strategy.

\subsection{Monadic Intermediate Languages}

Our transformation basically models the denotational semantics of Curry explicitly.
\citet{peytonjones1998bridging} have applied such an approach to design a common monadic intermediate language for Haskell and ML.
Although we do not use our monadic intermediate language to model two languages, the idea remains the same.

Transforming an effectful language into purely functional code using a monadic style has been used in the past to model functional languages in various proof assistant systems.
For instance, \citet{abel2005verifying} generate Agda code
that models Haskell's semantics via an explicit monadic effect.

More recently, a compilation scheme for the algebraic effect language
\textit{Eff} has been presented that translates Eff into monadic OCaml
\citep{karachalias2021efficient}.
However, Eff is an entirely different language than Curry
with different challenges.

\section{Conclusions and Future Work}
\label{sec:conclusions}

In this work, we developed an implementation of Curry
which supports various advanced features of functional logic languages
introduced in recent years.
In order to achieve a sufficient evaluation performance even when
combining lazy evaluation with non-determinism,
we adapted the recent method of memoized pull-tabbing from Curry compilers
that use imperative target languages to a functional language.
This evaluation strategy is modeled in a monadic style and
can be combined with an automatic transformation of Curry programs
into monadic Haskell programs.
We also extended this implementation to efficiently support
free variables from Curry and integrated other extensions,
like unification, functional patterns, encapsulated search, or fair search.
Due to the monadic structure of the target programs,
these extensions can be implemented by modifying the monad only
without changing the compilation scheme.
The results for small but typical benchmarks indicate a good performance
compared to other compilers.
Thus, we obtained a high-level, maintainable, and efficient
implementation of Curry which supports all language features
together with an operationally complete, fair search strategy.

For future work, there is a lot to do. Our determinism analysis and
optimization would of course be improved by user-given annotations in
the source language, i.e., enforcing that certain type class functions
cannot be non-deterministic. Additionally, using a strictness
analysis would also allow us to omit certain \code{share} usages and
omit laziness when it is not necessary.


\begin{thebibliography}{}

\bibitem[Abel et~al., 2005]{abel2005verifying}
{\sc Abel, A.}, {\sc Benke, M.}, {\sc Bove, A.}, {\sc Hughes, J.}, {\sc and}
  {\sc Norell, U.}
\newblock Verifying {Haskell} programs using constructive type theory.
\newblock In {\em Proceedings of the 2005 {{ACM SIGPLAN Workshop}} on
  {{Haskell}}} 2005, pp. 62--73, {New York, NY, USA}. {ACM Press}.

\bibitem[Albert et~al., 2005]{AlbertHanusHuchOliverVidal05}
{\sc Albert, E.}, {\sc Hanus, M.}, {\sc Huch, F.}, {\sc Oliver, J.}, {\sc and}
  {\sc Vidal, G.} 2005.
\newblock Operational semantics for declarative multi-paradigm languages.
\newblock {\em Journal of Symbolic Computation}, {\it 40}, 1, 795--829.

\bibitem[Alqaddoumi et~al., 2010]{AlqaddoumiAntoyFischerReck10}
{\sc Alqaddoumi, A.}, {\sc Antoy, S.}, {\sc Fischer, S.}, {\sc and} {\sc Reck,
  F.}
\newblock The pull-tab transformation.
\newblock In {\em Proc. of the Third International Workshop on Graph
  Computation Models} 2010, pp. 127--132, Enschede, The Netherlands. Published
  Online.
\newblock Available at http://gcm2010.imag.fr/pages/gcm2010-preproceedings.pdf.

\bibitem[Antoy, 1997]{Antoy97ALP}
{\sc Antoy, S.}
\newblock Optimal non-deterministic functional logic computations.
\newblock In {\em Proc. International Conference on Algebraic and Logic
  Programming (ALP'97)} 1997, pp. 16--30, Berlin, Heidelberg. Springer LNCS
  1298.

\bibitem[Antoy, 2001]{Antoy01PPDP}
{\sc Antoy, S.}
\newblock Constructor-based conditional narrowing.
\newblock In {\em Proc.\ of the 3rd International ACM SIGPLAN Conference on
  Principles and Practice of Declarative Programming (PPDP 2001)} 2001, pp.
  199--206, New York, NY, USA. ACM Press.

\bibitem[Antoy, 2011]{Antoy11ICLP}
{\sc Antoy, S.} 2011.
\newblock On the correctness of pull-tabbing.
\newblock {\em Theory and Practice of Logic Programming}, {\it 11}, 4-5,
  713--730.

\bibitem[Antoy and Bra{\ss}el, 2007]{AntoyBrassel07}
{\sc Antoy, S.} {\sc and} {\sc Bra{\ss}el, B.}
\newblock Computing with subspaces.
\newblock In {\em Proceedings of the 9th ACM SIGPLAN International Conference
  on Principles and Practice of Declarative Programming (PPDP'07)} 2007, pp.
  121--130. ACM Press.

\bibitem[Antoy et~al., 2000]{AntoyEchahedHanus00JACM}
{\sc Antoy, S.}, {\sc Echahed, R.}, {\sc and} {\sc Hanus, M.} 2000.
\newblock A needed narrowing strategy.
\newblock {\em Journal of the ACM}, {\it 47}, 4, 776--822.

\bibitem[Antoy and Hanus, 2000]{AntoyHanus00FROCOS}
{\sc Antoy, S.} {\sc and} {\sc Hanus, M.}
\newblock Compiling multi-paradigm declarative programs into {Prolog}.
\newblock In {\em Proc. International Workshop on Frontiers of Combining
  Systems (FroCoS'2000)} 2000, pp. 171--185, Berlin, Heidelberg. Springer LNCS
  1794.

\bibitem[Antoy and Hanus, 2002]{AntoyHanus02FLOPS}
{\sc Antoy, S.} {\sc and} {\sc Hanus, M.}
\newblock Functional logic design patterns.
\newblock In {\em Proc.\ of the 6th International Symposium on Functional and
  Logic Programming (FLOPS 2002)} 2002, pp. 67--87. Springer LNCS 2441.

\bibitem[Antoy and Hanus, 2005]{AntoyHanus05LOPSTR}
{\sc Antoy, S.} {\sc and} {\sc Hanus, M.}
\newblock Declarative programming with function patterns.
\newblock In {\em Proceedings of the International Symposium on Logic-based
  Program Synthesis and Transformation (LOPSTR'05)} 2005, pp. 6--22, Berlin,
  Heidelberg. Springer LNCS 3901.

\bibitem[Antoy and Hanus, 2006]{AntoyHanus06ICLP}
{\sc Antoy, S.} {\sc and} {\sc Hanus, M.}
\newblock Overlapping rules and logic variables in functional logic programs.
\newblock In {\em Proceedings of the 22nd International Conference on Logic
  Programming (ICLP 2006)} 2006, pp. 87--101, Berlin, Heidelberg. Springer LNCS
  4079.

\bibitem[Antoy and Hanus, 2009]{AntoyHanus09}
{\sc Antoy, S.} {\sc and} {\sc Hanus, M.}
\newblock Set functions for functional logic programming.
\newblock In {\em Proceedings of the 11th ACM SIGPLAN International Conference
  on Principles and Practice of Declarative Programming (PPDP'09)} 2009, pp.
  73--82, New York, NY, USA. ACM Press.

\bibitem[Antoy and Hanus, 2010]{AntoyHanus10CACM}
{\sc Antoy, S.} {\sc and} {\sc Hanus, M.} 2010.
\newblock Functional logic programming.
\newblock {\em Communications of the ACM}, {\it 53}, 4, 74--85.

\bibitem[Antoy et~al., 2020]{AntoyHanusJostLibby20}
{\sc Antoy, S.}, {\sc Hanus, M.}, {\sc Jost, A.}, {\sc and} {\sc Libby, S.}
\newblock {ICurry}.
\newblock In {\em Declarative Programming and Knowledge Management - Conference
  on Declarative Programming ({DECLARE} 2019)} 2020, pp. 286--307, Berlin,
  Heidelberg. Springer LNCS 12057.

\bibitem[Antoy and Jost, 2016]{AntoyJost16}
{\sc Antoy, S.} {\sc and} {\sc Jost, A.}
\newblock A new functional-logic compiler for {Curry}: {Sprite}.
\newblock In {\em Proceedings of the 26th International Symposium on
  Logic-Based Program Synthesis and Transformation (LOPSTR 2016)} 2016, pp.
  97--113, Berlin, Heidelberg. Springer LNCS 10184.

\bibitem[Baader and Nipkow, 1998]{BaaderNipkow98}
{\sc Baader, F.} {\sc and} {\sc Nipkow, T.} 1998.
\newblock {\em Term Rewriting and All That}.
\newblock Cambridge University Press, Cambridge, UK.

\bibitem[B{\"o}hm et~al., 2021]{BoehmHanusTeegen21}
{\sc B{\"o}hm, J.}, {\sc Hanus, M.}, {\sc and} {\sc Teegen, F.}
\newblock From non-determinism to goroutines: A fair implementation of {Curry}
  in {Go}.
\newblock In {\em Proc. of the 23rd International Symposium on Principles and
  Practice of Declarative Programming (PPDP 2021)} 2021, pp. 16:1--16:15, New
  York, NY, USA. ACM Press.

\bibitem[Bra{\ss}el et~al., 2004]{BrasselHanusHuch04JFLP}
{\sc Bra{\ss}el, B.}, {\sc Hanus, M.}, {\sc and} {\sc Huch, F.} 2004.
\newblock Encapsulating non-determinism in functional logic computations.
\newblock {\em Journal of Functional and Logic Programming}, {\it 2004}, 6.

\bibitem[Bra{\ss}el et~al., 2011]{BrasselHanusPeemoellerReck11}
{\sc Bra{\ss}el, B.}, {\sc Hanus, M.}, {\sc Peem{\"o}ller, B.}, {\sc and} {\sc
  Reck, F.}
\newblock {KiCS2}: A new compiler from {Curry} to {Haskell}.
\newblock In {\em Proc. of the 20th International Workshop on Functional and
  (Constraint) Logic Programming (WFLP 2011)} 2011, pp. 1--18, Berlin,
  Heidelberg. Springer LNCS 6816.

\bibitem[Bra{\ss}el and Huch, 2007]{BrasselHuch07}
{\sc Bra{\ss}el, B.} {\sc and} {\sc Huch, F.}
\newblock On a tighter integration of functional and logic programming.
\newblock In {\em Proc. APLAS 2007} 2007, pp. 122--138, Berlin, Heidelberg.
  Springer LNCS 4807.

\bibitem[Christiansen et~al., 2013]{ChristiansenHanusReckSeidel13PPDP}
{\sc Christiansen, J.}, {\sc Hanus, M.}, {\sc Reck, F.}, {\sc and} {\sc Seidel,
  D.}
\newblock A semantics for weakly encapsulated search in functional logic
  programs.
\newblock In {\em Proc. of the 15th International Symposium on Principle and
  Practice of Declarative Programming (PPDP'13)} 2013, pp. 49--60, New York,
  NY, USA. ACM Press.

\bibitem[de~Moura and Bj{\o}rner, 2008]{deMouraBjorner08}
{\sc de~Moura, L.} {\sc and} {\sc Bj{\o}rner, N.}
\newblock Z3: An efficient {SMT} solver.
\newblock In {\em Proc. of the 14th International Conference on Tools and
  Algorithms for the Construction and Analysis of Systems (TACAS 2008)} 2008,
  pp. 337--340. Springer LNCS 4963.

\bibitem[Eisenberg et~al., 2014]{Eisenberg14Closed}
{\sc Eisenberg, R.~A.}, {\sc Vytiniotis, D.}, {\sc Jones, S. L.~P.}, {\sc and}
  {\sc Weirich, S.}
\newblock Closed type families with overlapping equations.
\newblock In {\sc Jagannathan, S.} {\sc and} {\sc Sewell, P.}, editors, {\em
  The 41st Annual {ACM} {SIGPLAN-SIGACT} Symposium on Principles of Programming
  Languages, {POPL} '14, San Diego, CA, USA, January 20-21, 2014} 2014, pp.
  671--684. {ACM}.

\bibitem[Fischer et~al., 2011]{FischerKiselyovShan11}
{\sc Fischer, S.}, {\sc Kiselyov, O.}, {\sc and} {\sc Shan, C.} 2011.
\newblock Purely functional lazy nondeterministic programming.
\newblock {\em Journal of Functional programming}, {\it 21}, 4\&5, 413--465.

\bibitem[Gonz{\'a}lez-Moreno et~al., 1999]{GonzalezEtAl99}
{\sc Gonz{\'a}lez-Moreno, J.}, {\sc Hortal{\'a}-Gonz{\'a}lez, M.}, {\sc
  L{\'o}pez-Fraguas, F.}, {\sc and} {\sc Rodr{\'\i}guez-Artalejo, M.} 1999.
\newblock An approach to declarative programming based on a rewriting logic.
\newblock {\em Journal of Logic Programming}, {\it 40}, 47--87.

\bibitem[Hanus, 2011]{Hanus11ICLP}
{\sc Hanus, M.}
\newblock Declarative processing of semistructured web data.
\newblock In {\em Technical Communications of the 27th International Conference
  on Logic Programming} 2011, volume~11, pp. 198--208. Leibniz International
  Proceedings in Informatics (LIPIcs).

\bibitem[Hanus, 2012]{Hanus12ICLP}
{\sc Hanus, M.}
\newblock Improving lazy non-deterministic computations by demand analysis.
\newblock In {\em Technical Communications of the 28th International Conference
  on Logic Programming} 2012, volume~17, pp. 130--143, Dagstuhl, Germany.
  Leibniz International Proceedings in Informatics (LIPIcs).

\bibitem[Hanus, 2013]{Hanus13}
{\sc Hanus, M.}
\newblock Functional logic programming: From theory to {Curry}.
\newblock In {\em Programming Logics - Essays in Memory of Harald Ganzinger}
  2013, pp. 123--168, Berlin, Heidelberg. Springer LNCS 7797.

\bibitem[Hanus, 2024]{Hanus24LOPSTR}
{\sc Hanus, M.}
\newblock Improving logic programs by adding functions.
\newblock In {\em Proceedings of the 34th International Symposium on
  Logic-Based Program Synthesis and Transformation (LOPSTR 2024)} 2024, pp.
  27--44. Springer LNCS 14919.

\bibitem[Hanus et~al., 2025]{Hanus25PAKCS}
{\sc Hanus, M.}, {\sc Antoy, S.}, {\sc Bra{\ss}el, B.}, {\sc Engelke, M.}, {\sc
  H{\"o}ppner, K.}, {\sc Koj, J.}, {\sc Niederau, P.}, {\sc Sadre, R.}, {\sc
  Steiner, F.}, {\sc and} {\sc Teegen, F.} 2025.
\newblock {PAKCS}: The {Portland} {Aachen} {Kiel} {Curry} {System}.
\newblock Available at \url{https://www.curry-lang.org/pakcs/}.

\bibitem[Hanus et~al., 2012]{HanusPeemoellerReck12ATPS}
{\sc Hanus, M.}, {\sc Peem{\"o}ller, B.}, {\sc and} {\sc Reck, F.}
\newblock Search strategies for functional logic programming.
\newblock In {\em Proc. of the 5th Working Conference on Programming Languages
  (ATPS'12)} 2012, pp. 61--74, Bonn. Springer LNI 199.

\bibitem[Hanus et~al., 2022]{HanusProttTeegen22}
{\sc Hanus, M.}, {\sc Prott, K.-O.}, {\sc and} {\sc Teegen, F.}
\newblock A monadic implementation of functional logic programs.
\newblock In {\em Proc. of the 24th International Symposium on Principles and
  Practice of Declarative Programming (PPDP 2022)} 2022, pp. 1:1--1:15. ACM
  Press.

\bibitem[Hanus and Skrlac, 2014]{HanusSkrlac14}
{\sc Hanus, M.} {\sc and} {\sc Skrlac, F.}
\newblock A modular and generic analysis server system for functional logic
  programs.
\newblock In {\em Proc. of the ACM SIGPLAN 2014 Workshop on Partial Evaluation
  and Program Manipulation (PEPM'14)} 2014, pp. 181--188. ACM Press.

\bibitem[Hanus and Teegen, 2020]{HanusTeegen20}
{\sc Hanus, M.} {\sc and} {\sc Teegen, F.}
\newblock Adding \texttt{Data} to {Curry}.
\newblock In {\em Declarative Programming and Knowledge Management - Conference
  on Declarative Programming ({DECLARE} 2019)} 2020, pp. 230--246. Springer
  LNCS 12057.

\bibitem[Hanus and Teegen, 2021]{HanusTeegen21}
{\sc Hanus, M.} {\sc and} {\sc Teegen, F.}
\newblock Memoized pull-tabbing for functional logic programming.
\newblock In {\em Proc. of the 28th International Workshop on Functional and
  (Constraint) Logic Programming (WFLP 2020)} 2021, pp. 57--73, Berlin,
  Heidelberg. Springer LNCS 12560.

\bibitem[Hanus~{(ed.)}, 2016]{Hanus16Curry}
{\sc Hanus~{(ed.)}, M.} 2016.
\newblock Curry: An integrated functional logic language (vers.\ 0.9.0).
\newblock Available at \url{https://www.curry-lang.org}.

\bibitem[Huet and L{\'e}vy, 1991]{HuetLevy91}
{\sc Huet, G.} {\sc and} {\sc L{\'e}vy, J.-J.}
\newblock Computations in orthogonal rewriting systems.
\newblock In {\sc Lassez, J.-L.} {\sc and} {\sc Plotkin, G.}, editors, {\em
  Computational Logic: Essays in Honor of Alan Robinson} 1991, pp. 395--443.
  MIT Press, Cambridge, Massachusetts.

\bibitem[Hussmann, 1992]{Hussmann92}
{\sc Hussmann, H.} 1992.
\newblock Nondeterministic algebraic specifications and nonconfluent term
  rewriting.
\newblock {\em Journal of Logic Programming}, {\it 12}, 237--255.

\bibitem[Johnsson, 1985]{Johnsson85}
{\sc Johnsson, T.}
\newblock Lambda lifting: Transforming programs to recursive functions.
\newblock In {\em Functional Programming Languages and Computer Architecture}
  1985, pp. 190--203, Berlin, Heidelberg. Springer LNCS 201.

\bibitem[Karachalias et~al., 2021]{karachalias2021efficient}
{\sc Karachalias, G.}, {\sc Koprivec, F.}, {\sc Pretnar, M.}, {\sc and} {\sc
  Schrijvers, T.} 2021.
\newblock Efficient compilation of algebraic effect handlers.
\newblock {\em Proc. ACM Program. Lang.}, {\it 5}, OOPSLA.

\bibitem[Lloyd, 1987]{Lloyd87}
{\sc Lloyd, J.} 1987.
\newblock {\em Foundations of Logic Programming}.
\newblock Springer, second, extended edition, Berlin, Heidelberg.

\bibitem[L\'opez-Fraguas and S\'anchez-Hern\'andez,
  1999]{Lopez-FraguasSanchez-Hernandez99}
{\sc L\'opez-Fraguas, F.} {\sc and} {\sc S\'anchez-Hern\'andez, J.}
\newblock {TOY}: A multiparadigm declarative system.
\newblock In {\em Proc. of RTA'99} 1999, pp. 244--247, Berlin, Heidelberg.
  Springer LNCS 1631.

\bibitem[M{oreno-Navarro} et~al., 1990]{MorenoEtAl90ALP}
{\sc M{oreno-Navarro}, J.}, {\sc Kuchen, H.}, {\sc Loogen, R.}, {\sc and} {\sc
  R{odr{\'\i}guez-Artalejo}, M.}
\newblock Lazy narrowing in a graph machine.
\newblock In {\em Proc. Second International Conference on Algebraic and Logic
  Programming} 1990, pp. 298--317. Springer LNCS 463.

\bibitem[Perry, 2005]{Perry2005Thesis}
{\sc Perry, N.} 2005.
\newblock {\em The Implementation of Practical Functional Programming
  Languages}.
\newblock PhD thesis, University of London.

\bibitem[Petricek, 2012]{Petricek12}
{\sc Petricek, T.} 2012.
\newblock Evaluation strategies for monadic computations.
\newblock {\em Electronic Proceedings in Theoretical Computer Science}, {\it
  76}, 68--89.

\bibitem[Peyton~Jones, 2003]{PeytonJones03Haskell}
{\sc Peyton~Jones, S.}, editor 2003.
\newblock {\em Haskell 98 Language and Libraries---The Revised Report}.
\newblock Cambridge University Press, Cambridge, UK.

\bibitem[Peyton~Jones et~al., 1996]{PeytonJonesGordonFinne96POPL}
{\sc Peyton~Jones, S.}, {\sc Gordon, A.}, {\sc and} {\sc Finne, S.}
\newblock Concurrent {Haskell}.
\newblock In {\em Proc. 23rd ACM Symposium on Principles of Programming
  Languages (POPL'96)} 1996, pp. 295--308. ACM Press.

\bibitem[Peyton~Jones et~al., 2000]{PeytonJones2000Finalize}
{\sc Peyton~Jones, S.}, {\sc Marlow, S.}, {\sc and} {\sc Elliott, C.}
\newblock Stretching the storage manager: Weak pointers and stable names in
  {Haskell}.
\newblock In {\sc Koopman, P.} {\sc and} {\sc Clack, C.}, editors, {\em
  Implementation of Functional Languages} 2000, pp. 37--58, Berlin, Heidelberg.
  Springer Berlin Heidelberg.

\bibitem[Peyton~Jones et~al., 1998]{peytonjones1998bridging}
{\sc Peyton~Jones, S.}, {\sc Shields, M.}, {\sc Launchbury, J.}, {\sc and} {\sc
  Tolmach, A.}
\newblock Bridging the gulf: {{A}} common intermediate language for {{ML}} and
  {Haskell}.
\newblock In {\em Proceedings of the 25th {{ACM SIGPLAN}}-{{SIGACT}} Symposium
  on Principles of Programming Languages} 1998, {{POPL}} '98, pp. 49--61, {New
  York, NY, USA}. {Association for Computing Machinery}.

\bibitem[Prott et~al., 2023]{Prott23Embedding}
{\sc Prott, K.}, {\sc Teegen, F.}, {\sc and} {\sc Christiansen, J.}
\newblock Embedding functional logic programming in {Haskell} via a compiler
  plugin.
\newblock In {\sc Hanus, M.} {\sc and} {\sc Inclezan, D.}, editors, {\em
  Practical Aspects of Declarative Languages - 25th International Symposium,
  {PADL} 2023, Boston, MA, USA, January 16-17, 2023, Proceedings} 2023, volume
  13880 of {\em Lecture Notes in Computer Science}, pp. 37--55. Springer.

\bibitem[Reddy, 1985]{Reddy85}
{\sc Reddy, U.}
\newblock Narrowing as the operational semantics of functional languages.
\newblock In {\em Proc. IEEE International Symposium on Logic Programming}
  1985, pp. 138--151, Boston. IEEE Computer Society.

\bibitem[Robinson, 1965]{Robinson65}
{\sc Robinson, J.} 1965.
\newblock A machine-oriented logic based on the resolution principle.
\newblock {\em Journal of the ACM}, {\it 12}, 1, 23--41.

\bibitem[Slagle, 1974]{Slagle74}
{\sc Slagle, J.} 1974.
\newblock Automated theorem-proving for theories with simplifiers,
  commutativity, and associativity.
\newblock {\em Journal of the ACM}, {\it 21}, 4, 622--642.

\bibitem[Teegen et~al., 2021]{teegen21inversion}
{\sc Teegen, F.}, {\sc Prott, K.-O.}, {\sc and} {\sc Bunkenburg, N.}
\newblock {{Haskell}}{$^{-1}$}: Automatic function inversion in {Haskell}.
\newblock In {\em Proceedings of the 14th {{ACM SIGPLAN}} International
  Symposium on Haskell} 2021, Haskell 2021, pp. 41--55, {New York, NY, USA}.
  {Association for Computing Machinery}.

\bibitem[Wadler, 1985]{Wadler85}
{\sc Wadler, P.}
\newblock How to replace failure by a list of successes: {A} method for
  exception handling, backtracking, and pattern matching in lazy functional
  languages.
\newblock In {\em Conference on Functional Programming and Computer
  Architecture (FPCA'85)} 1985, pp. 113--128, Berlin, Heidelberg. Springer LNCS
  201.

\bibitem[Wadler, 1990]{Wadler90LFP}
{\sc Wadler, P.}
\newblock Comprehending monads.
\newblock In {\em Proc. 1990 {ACM} Conference on {LISP} and Functional
  Programming} 1990, pp. 61--78, New York, NY, USA. ACM.

\bibitem[Wadler, 1997]{Wadler97}
{\sc Wadler, P.} 1997.
\newblock How to declare an imperative.
\newblock {\em ACM Computing Surveys}, {\it 29}, 3, 240--263.

\end{thebibliography}

\end{document}